\documentclass[8pt,twocolumn,showpacs,preprintnumbers,amsmath,amssymb,prb,superscriptaddress]{revtex4}
\usepackage{txfonts}
\usepackage{amssymb}

\usepackage{graphicx}
\usepackage{dcolumn}
\usepackage{bm}

\begin{document}

\title{Using spin bias to manipulate and measure spin in quantum dots}
\author{Hai-Zhou Lu}
\affiliation{Department of Physics, and Centre of Theoretical and Computational Physics,
The University of Hong Kong, Pokfulam Road, Hong Kong, China}
\author{Shun-Qing Shen}
\affiliation{Department of Physics, and Centre of Theoretical and Computational Physics,
The University of Hong Kong, Pokfulam Road, Hong Kong, China}

\begin{abstract}
A double quantum dot coupled to electrodes with spin-dependent splitting of
chemical potentials (spin bias) is investigated theoretically by means of
the nonequilibrium Kyldysh Green's function formalism. By applying a large
spin bias, the quantum spin in a quantum dot ( dot 1 ) can be manipulated
in a fully electrical manner. To noninvasively monitor the manipulation of
the quantum spin in dot 1, it is proposed that the second quantum dot
( dot 2 ) is weakly coupled to dot 1. In the presence of the exchange
interaction between the two dots, the polarized spin in dot 1 behaves
like an effective magnetic field and weakly polarizes the spin in the nearby
quantum dot 2. By applying a very small spin bias to dot 2, the
spin-dependent transport through dot 2 can be probed, allowing the spin
polarization in dot 1 to be identified nondestructively. These two steps
form a complete scheme to manipulate a trapped spin while permitting this
manipulation to be monitored in the double-dot system using pure electric
approaches.
\end{abstract}

\pacs{85.75.-d, 73.21.La, 72.25.Hg}
\date{\today }
\maketitle

\section{Introduction}

The manipulation and measurement of single electron spin in a quantum
dot is the basis toward scalable spin-based quantum information
processing.\cite{Loss1998} The preparation and readout of a single
spin in a quantum dot have been demonstrated using photoluminescence
polarization\cite{Bracker2005,Ebbens2005} and polarization-dependent
absorption.\cite{Stievater2002,Li2004,Hogele2005} The rapid progress
of the charging sensing technique\cite{Sprinzak2002,Elzerman2003}
makes it possible to control the number of electrons inside quantum
dots precisely down to a few electrons,\cite{Hanson2007} allowing an
individual electron spin to be manipulated with the help of stationary
and oscillating electromagnetic field, \cite{Koppens2006,Nowack2007}
and the readout by various spin-to-charge conversion techniques,
such as single-shot readout using energy\cite{Elzerman2004} or
tunneling rate difference\cite{Hanson2005} and Pauli spin
blockade.\cite{Koppens2006} However, most of the detection
techniques destroy the originally prepared spin state. All the
electric approaches remove the trapped spin from its host dot, and
most of optical measurements drive the spin polarized states to
other states. A noninvasive detection method is needed because a
complete control process requires the manipulated spin to be
monitored and not be affected by the monitoring. Recently, using
off-resonant picosecond-scale optical pulses and time-resolved Kerr
rotation spectroscopy, single electron spin in a quantum dot is
nondestructively measured,\cite{Berezovsky2006} which further leads
to better manipulation of the spin.\cite{Berezovsky2008} Besides, by
using energy-dependent single-shot readout followed by immediate
restoring of the spin back to dot within a time shorter than the
spin relaxation time, up
to 90\% of the original spin states can be retained after the measurement.%
\cite{Meunier2006} Still, all the techniques require precisely controlled
gating, electromagnetic or optical field, and their time scale has to be
within the spin coherence time considering various decoherence mechanisms
in host materials.

A hint from the charging sensing technique\cite{Sprinzak2002,Elzerman2003}
is that it makes use of a nearby quantum point contact to noninvasively
measure the electron number in the quantum dot, which avoids destroying the
electron occupation in the quantum dot by direct transport measurement. From
the point of view of spin-based quantum information processing, it would be
highly desirable to design a similar device with integrated ability to
manipulate a trapped single spin while permitting the manipulation to be
read out nondestructively in quantum dot. Most importantly, a pure electric
approach (in the absence of magnetic or optical field) is particularly
appealing for a large-scale integration.

A possible direction for this effort points to the spin injection technique, i.e.,
generating a nonequilibrium spin accumulation in nonmagnetic
(paramagnetic) materials, which could induce a spin-dependent
splitting of chemical potentials or spin bias in the injected
materials. Spin injection has been demonstrated using several of
electric and optical approaches. One of the effective methods is to
inject spin-polarized charge current directly from ferromagnetic to
nonmagnetic materials.\cite{Zutic2004} The unequal density of states
for majority and minority spins at the Fermi level of the
ferromagnet will induce a spin accumulation in the nonmagnetic
material and split the chemical potentials for two spin components.
The materials and geometries used in this method include
metals,\cite{Johnson1985,Johnson1988d,Valenzuela2006}
metal/barrier/semiconductors,
\cite{Zhu2001,Motsnyi2002,Hanbicki2002} and ferromagnetic/normal
semiconductors.\cite{Fiederling1999, Ohno1999} In the last few
years, spin injection has also been shown by means of the spin Hall
effect\cite{Kato2004sci,Wunderlich2005} and the incidence of linearly
or circularly polarized light into a two-dimensional electron gas with
spin-orbital coupling.\cite{Ganichev2006,Cui2007,Li2006} Notice that
many of the intensively investigated spin-injected nonmagnetic
materials are also widely used to fabricate electrodes probing
semiconductor\cite{Sohn1997} and single-molecule quantum
dots.\cite{Park2000} Therefore, it is interesting to investigate the
polarization and detection of electron spin in quantum dot systems
using electrodes with spin bias. Experimentally, spin injection into
all-semiconductor quantum dots has been reported, from (Ga,Mn)As to InAs quantum dots (QDs)
\cite{Chye2002}, and from BeMnZnSe to a single CdSe/ZnSe QD, both
combined with a spin-light-emitting-diode to detect spin polarization. Furthermore, several
theoretical works addressed the transport through mesoscopic
systems in the presence of spin-splitting of chemical potentials.
\cite{Veillette2004,Zhang2003,Wang2004,li2007}

\begin{figure}[h]
\centering \includegraphics[width=0.35\textwidth]{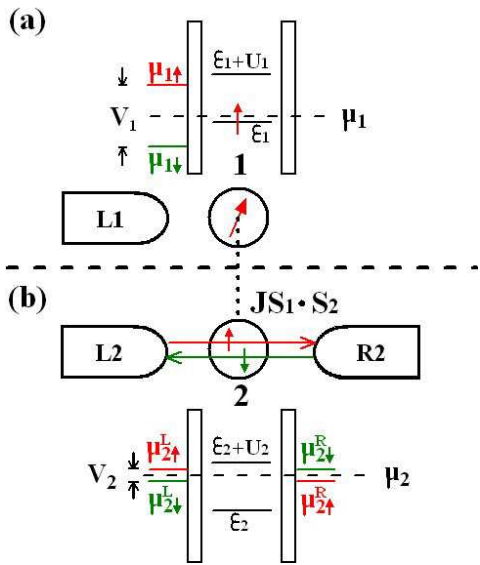}
\caption{Schematic of our double-dot
system, in which each dot is attached to its own electrodes with
spin-dependent splitting of chemical potentials (spin biases). (a) A
large spin bias $V_1$ is applied to manipulate the quantum spin in
dot 1. The energy zero point is set at $ \epsilon_1+U_1/2$. The
spin bias $V_1$ induces a splitting of the Fermi levels for
$\uparrow$ and $\downarrow$ electrons in lead L1 so that $
\mu_{1\uparrow/\downarrow}= \mu_1\pm V_1/2$, where $ \mu_1$ is the
middle point of the Fermi levels. (b) Due to the exchange
interaction between the two dots, the polarized spin in dot 1
behaves like an effective magnetic field and weakly polarizes the
spin in dot 2. A very small spin bias $V_2$ is applied to the
dot 2 to probe its spin-dependent transport, allowing the spin
polarization in dot 1 to be read out nondestructively. $ \mu_2$
is the equilibrium Fermi level for both leads L2 and R2. The spin
bias $V_2$ induces a splitting of the Fermi levels for
$\uparrow$ ($\downarrow$) electrons in the
leads of dot 2, so that $ \mu_{2\uparrow/\downarrow}^{L}= \mu%
_2\pm V_2/2$ and $ \mu_{2\uparrow/\downarrow}^{R}= \mu_2\mp V_2/2$.
} \label{fig:doubledot}
\end{figure}

Motivated by these experimental and theoretical progresses, we
propose a scheme to realize the control and detection of quantum
spin in semiconductor quantum dot by using spin bias or spin
current. Our setup consists of a double-quantum-dot system connected
to electrodes as shown in Fig. \ref{fig:doubledot}. A quantum spin
state can be generated and maintained in dot 1 when applying a
spin bias $V_{1}$ on the electrode coupling the quantum dot 1 [Fig.
\ref{fig:doubledot}(a)]. In the presence of exchange interaction
between the two dots, the polarized spin in dot 1 behaves like
an effective magnetic field and breaks the spin symmetry in dot
2. As a result, it will induce a charge current when a small spin
bias $V_{2}$ is applied or a spin current flows through dot 2,
allowing the spin polarization in dot 1 to be identified [Fig.
\ref{fig:doubledot} (b)]. If the interdot exchange interaction is
much smaller than $V_{1}$, the measurement can be viewed as
nondestructive. These two steps form a complete scheme of
manipulating and measuring the quantum spin state of a trapped
electron in one dot of a double-dot system using purely electric
means. Our proposal is based on steady-state evaluation; no
ultrafast optical or electrostatic operation is needed. We argue
that it is robust once the magnitude of $V_{1}$ energetically
overwhelms those of decoherence mechanisms, such as hyperfine
interaction with nuclear spins of host materials, or spin-orbital
coupling. It is worth stressing that as the spin injection
techniques of various means are still under extensive investigations
and progress, in the present work, we focus only on the physical
consequences of spin bias or spin current, and ignore the approaches
to generate the spin bias at the current stage.

The present paper is organized as follows: In Sec.
\ref{sec:proposal}, we present the general scheme and description of the
manipulation and detection of a single spin using the idea of spin
bias, by comparing with the known charge sensing technique. The
model Hamiltonian is introduced. The experimental feasibility of our
scheme is discussed based on recent experimental availability. In
Sec. \ref{Sec:dot 1}, the manipulation of quantum spin in dot 1
is addressed. The stability diagrams of electron number and spin
polarization in dot 1 are presented. In Sec.
\ref{Sec:detection}, the measurement of the spin polarization in the
dot 1 using the spin-dependent transport through dot 2 is
discussed. A detailed analytical calculation and the numerical results
are presented. We focus on the charge current induced by the spin
bias or spin conductance through dot 2 and its relation with
the spin polarization in dot 1. In Sec. \ref{Sec:summary}, a
summary is presented. Finally, the detailed calculations of spin
conductance and Green's functions are presented in Appendices A and B for
reference.

\section{\label{sec:proposal}General Proposal}

\subsection{\label{sec:scheme}Spin sensing scheme}

A typical setup for charge sensing technique (sketched in the second
column of Table \ref{tab:summary}) consists of a quantum dot which
hosts the electrons to be manipulated and a nearby quantum point
contact.\cite{Sprinzak2002} The electron number inside the
host dot can be determined by directly probing the transport through
dot; however, this approach implies removing electrons from dot, destroying the original occupation.
Alternatively, due to Coulomb
repulsion, the transport through the quantum point contact is found
to be very sensitive to the electron number in dot and thus can be
used to determine the electron number. Most importantly, this approach is noninvasive
because it does not change the electron number in dot.

Our scheme combining spin manipulation and detection in one setup employs a
similar idea. We apply the electron spin to play the role of the electron
charge in the device. As shown in Fig. \ref{fig:doubledot}, the setup is
composed of two quantum dots. dot 1 is the host for the manipulated
spin. dot 2 is used to nondestructively detect the spin in dot 1,
analogous to the function of quantum point contact in charge sensing.
Each dot is coupled independently to its nonmagnetic leads. The Fermi levels
of the leads are spin dependent and can be split by the spin bias. Without
loss of generality, we consider only one spin-degenerate energy level in
each dot, denoted as $\epsilon _{i}$ for a dot $i$ ($i=1,2$). The two dots
are assumed to be coupled weakly to each other via the Heisenberg exchange
coupling with strength $J$.

To manipulate the spin in dot 1, we apply a spin bias $V_{1}$ to lead L1 attached to dot 1.
To retain the electrons in dot 1, we consider
only one reservoir to avoid electron transport in the usual two-reservoir
case. We denote by $\mu _{1\sigma }$ the Fermi level for $\sigma $ electrons
in lead L1. The spin bias $V_{1}$ induces a splitting of the Fermi
levels for spin $\uparrow $ and $\downarrow $ electrons in lead L1 so
that
\begin{eqnarray*}
\mu _{1\uparrow } &=&\mu _{1}+V_{1}/2, \\
\mu _{1\downarrow } &=&\mu _{1}-V_{1}/2,
\end{eqnarray*}
where $\mu _{1}$ is the middle point of the Fermi levels,
\begin{equation*}
\mu _{1}=\frac{1}{2}(\mu _{1\uparrow }+\mu _{1\downarrow }),
\end{equation*}
which can be tuned with respect to $\epsilon _{1}$ by using the gate voltage.
The energy zero point is set at $\epsilon _{1}+U_{1}/2$. When the
dot 1 is coupled to its reservoir, the electron number inside it at
low temperatures will be determined by the relative location between
dot level $\epsilon _{1}$ and the Fermi levels of the reservoir.
In the absence of spin-dependent splitting of chemical potentials
($\mu _{1\uparrow }=\mu _{1\downarrow }$), it is known that dot
will be filled or empty when the Fermi levels are well above or
below dot level. Similarly, in the presence of the spin-resolved
Fermi levels, the filling of spin $\uparrow $ ($\downarrow $)
electron in dot 1 is determined independently by the relative
location between $\epsilon _{1}$ and $\mu _{1\uparrow }$ ($\mu
_{1\downarrow }$). The simplest situation, shown in Fig.
\ref{fig:doubledot}, is that when $\mu _{1\downarrow }<\epsilon _{1}<\mu
_{1\uparrow }$, only $\uparrow $ electron is energetically allowed
to stay in dot; therefore, the electron spin in dot is
$\uparrow $ polarized. When considering the on-site Coulomb
repulsion, detailed calculations are required; these are presented in Sec.
\ref{Sec:dot 1}.

In the presence of exchange interaction, once the electron in dot 1
is polarized by $V_{1}$, it will act approximately as an effective magnetic
field and weakly polarize the electron in the nearby dot 2. The polarized
spin in dot 2 will, in turn, act as an effective magnetic field to dot
1 and will influence the polarization of dot 1. If there is no extra
mechanism to break the spin symmetry, this self-consistent process repeats
until $\langle s_{1}^{z}\rangle $ and $\langle s_{2}^{z}\rangle $ approach
zero, where $\langle s_{i}^{z}\rangle \equiv \frac{1}{2}(\langle
n_{i\uparrow }\rangle -\langle n_{i\downarrow }\rangle )$ is the spin
polarization in dot $i$ and $\langle n_{i\sigma }\rangle $ is the
electron occupation for spin-$\sigma $ electron in dot $i$. However, in
the limit
\begin{equation}
V_{1}>>J\langle s_{2}^{z}\rangle ,  \label{V12}
\end{equation}%
the spin polarization in dot 1 is dominantly determined by the spin bias
$V_{1}$ and is hardly affected by the effective magnetic field generated by
dot 2. Once the spin is polarized in dot 1, it imposes a selection
rule to the spin orientation of electrons that can tunnel through dot 2
because the ground state of the two spins that are singly occupied in each dot tends
to form a fixed alignment in the presence of the exchange interaction.
Therefore, by measuring the spin polarization of the current that can flow
through dot 2, the spin orientation of dot 1 can be read out. Because
of Eq. (\ref{V12}), this measurement can be viewed as nondestructive.

To measure the spin-polarization of the current that can flow through the
dot 2, a small spin bias $V_2$ is applied to the electrodes attached to the
dot 2 so that the Fermi levels split in the left lead,
\begin{equation*}
\mu _{2\uparrow /\downarrow }^{L}=\mu _{2}\pm V_{2}/2,
\end{equation*}
and in the right lead,
\begin{equation*}
\mu _{2\uparrow /\downarrow }^{R}=\mu _{2}\mp V_{2}/2,
\end{equation*}
where $\mu _{2}$ is the Fermi level when there is no spin and charge
biases. Notice that in the presence of the pure spin bias $V_{2}$,
electrons of spin $\uparrow $ and $\downarrow $ will flow along
opposite directions through dot 2. In our definition, the
spin-$\uparrow$(-$\downarrow$) current can only flow from the left
(right) to the right (left). If the up-down symmetry of the electron
spin in dot is not broken, electric currents of spin $\uparrow $
and $\downarrow $ will cancel each other. Consequently a pure
spin current is formed.\cite{Sun2003} However, if the spin symmetry
of dot 2 is broken, the pure spin bias $V_{2}$ applied to the
dot 2 will generate a net electric current. In this way, measuring
the direction of current through dot 2 in the presence of $V_2$
is enough to determine the spin polarization of the current through dot 2 and the spin orientation in dot 1.
Furthermore, to avoid the
polarization of the spin in dot 2 affected by $V_2$, we demand
that
\begin{equation}  \label{V2<J}
J\langle s_{1}^{z}\rangle >>V_{2}.
\end{equation}

\subsection{\label{sec:hamiltonian}Model for coupled double dot}

The Hamiltonian for the whole double-dot consists of three parts,
\begin{equation}
H=H_{1}+H_{2}+V_{12},  \label{H12}
\end{equation}
where $H_{i}$ is the Hamiltonian for a dot $i$ and its leads alone,
described by the Anderson model,\cite{Anderson1961}
\begin{eqnarray}
H_{i} &=&\sum_{\sigma }\epsilon _{i}n_{i\sigma }+U_{i}n_{i\uparrow
}n_{i\downarrow }+\sum_{k,\sigma }\epsilon _{k\alpha \sigma }c_{k\alpha
\sigma }^{\dag }c_{k\alpha \sigma }  \notag  \label{onedotH} \\
&&+\sum_{k,\sigma }(V_{k\alpha \sigma }c_{k\alpha \sigma }^{\dag }d_{i\sigma
}+h.c.),
\end{eqnarray}%
where $d_{i\sigma }^{\dag }(d_{i\sigma })$ represents the creation
(annihilation) operator for the discrete state with  energy $\epsilon _{i}
$ and spin $\sigma $($\in \{\uparrow ,\downarrow \}$) in dot $i$, the
number operator $n_{i\sigma }=d_{i\sigma }^{\dag }d_{i\sigma }$, and $U_{i}$
is the intra-dot Coulomb repulsion. $c_{k\alpha \sigma }^{\dag }(c_{k\alpha
\sigma })$ is the creation (annihilation) operator for a continuous state in
the $\alpha $ lead (reservoir) with energy $\epsilon _{k\alpha \sigma }$ and
spin $\sigma $. The tunneling matrix element $V_{k\alpha \sigma }$ is
assumed to be independent of $k$ in the following calculations.

$V_{12}$ in Eq. (\ref{H12}) stands for the interaction between the two
dots. Basically, there are three forms of interactions:

(1) The tunneling coupling ($t_{c}d_{1\sigma }^{\dag }d_{2\sigma
}+h.c$).\cite{Holleitner2001} When the two dots are very closely
located electrons are allowed to tunnel between the two dots
directly. It should be avoided if one intends to perform a
noninvasive measurement. We will discuss how to prevent it in Sec.
\ref{sec:validity}.

(2) The capacitive coupling $U^{\prime }n_{1}n_{2}$.\cite{Chan2002}
The Coulomb repulsive interaction always exists when the two dots
are closely located, but well separated. The occupancy of electrons
in one dot will affect the charge transport in the other dot. This is
the microscopic mechanism for the charge sensing technique. However
it is not spin-resolved, and we shall ignore it in our calculation.
We will explain in Sec. \ref{Sec:dot 2} that neglecting this term
brings no qualitatively change.

(3) The Heisenberg exchange coupling,
\begin{eqnarray}  \label{V_12}
V_{12}=J\mathbf{s}_{1}\cdot \mathbf{s}_{2}
\end{eqnarray}
where $\mathbf{s}_{i}=\frac{1}{2}\sum_{\sigma ,\sigma ^{\prime
}}d_{i\sigma }^{\dag }\hat{\sigma}_{\sigma \sigma ^{\prime
}}d_{i\sigma ^{\prime }}$ and $\hat{\sigma}=(\sigma _{x},\sigma
_{y},\sigma _{z})$ are the Pauli matrices.\cite{Chen2004} It does
not change the occupation number of each dot; but it affects the states
of electron spin, in particular, when the electrons are spin resolved. To
simplify our calculation, we consider only the exchange coupling.

\begin{figure}[h]
\centering \includegraphics[width=0.35\textwidth]{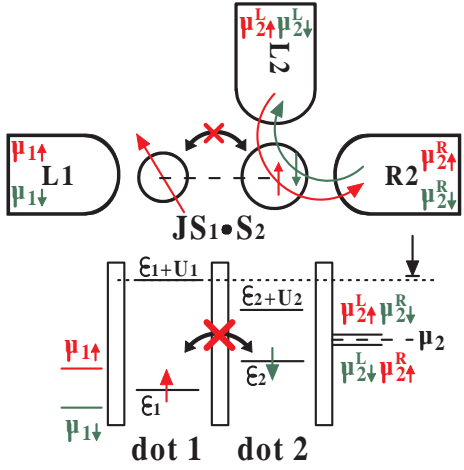}\newline
\caption{Schematic and energy
configuration of the low-energy effective double-dot model. The
interdot first-order direct tunneling is suppressed due to careful
design of all characteristic energies. The second-order virtual
hopping induces a low-energy Heisenberg exchange interaction with a
considerable positive exchange strength $J$. $ \mu_{2\sigma}^{
\alpha}$ is the Fermi level for spin $\sigma$($\in
\{\uparrow,\downarrow\}$) electrons in the $\alpha$ ($\in$\{L2,
R2\}) lead of dot 2.
$\mu_2=\frac{1}{2}(\mu_{2\uparrow}^{L/R}+\mu_{2\downarrow}^{L/R})$
is the shared middle point of Fermi levels for both leads L2 and
R2.} \label{fig:Chang}
\end{figure}

\subsection{\label{sec:validity}Experimental feasibility}

We have to point out that the model in Eqs. (\ref{H12}) and (\ref{V_12})
with considerable $J$ is valid in experiments only as a low-energy effective
Hamiltonian when direct electron hopping is quenched between two
tunneling-coupled quantum dots (coupling constant $t_{c}$ can be as large as
hundreds of $\mu $eV).\cite{Chen2004,Petta2005} To suppress the direct
tunneling and to employ only the low-energy spin dynamics between the two
dots, the energy configuration of the double dot should be carefully
designed, as shown in Fig. \ref{fig:Chang}. The level energies of the two
dots can be adjusted with respect to each other by tuning the gate voltages, while
the charging energies of the two dots can be customized by engineering the dot sizes.%
\cite{Chen2004} We choose an energy configuration such that $\epsilon
_{2}-\epsilon _{1}>|t_{c}|$ and $(\epsilon _{1}+U_{1})-(\epsilon
_{2}+U_{2})>|t_{c}|$. For dot 1, $\mu _{1}$ and $V_{1}$ are restricted
to the singly occupied regimes shown in Fig. \ref{fig:9 configurations}.
Moreover, since $\epsilon _{1}<\epsilon _{2}$, the electron in dot 1,
which resides on $\epsilon _{1}$, cannot hop to dot 2 no matter whether
dot 2 is occupied or not. For dot 2, the scanning range of the
middle point of the Fermi level $\mu _{2}$ is well restricted, $<\epsilon
_{1}+U_{1}$ (as shown by dotted line in Fig. \ref{fig:Chang}). As a
result, although double occupation in dot 2 is allowed when $\mu
_{2}>\epsilon _{2}+U_{2}$, the electrons in dot 2 still cannot hop to
dot 1 because they can not acquire enough energy from $\mu _{2\sigma
}^{\alpha }$ to conquer the charging energy $U_{1}$ for double occupation in
dot 1. The above conditions assure that there is no direct hopping
between the two dots. However, due to the uncertainty principle, within a
time $\sim \hbar /U_{i}$ or $\sim \hbar /|\epsilon _{1}-\epsilon _{2}|$, an
electron in dot 1 (2) can still possibly hop to dot 2 (1) and back.
Because of the Pauli exclusion principle, this kind of second-order virtual
hopping favors the anti-parallel alignment of two spins in the double dot.
As a result, the low-energy correlation between the two dots is well
described by the Heisenberg Hamiltonian with a positive exchange coupling
strength $J$. Using the simplest Hubbard model estimate $J=\frac{4t_{c}^{2}}{%
U}$, $J$ can be as large as $0.09$ meV when $t_{c}=0.15$ meV and $U\sim 1$%
meV.\cite{Chen2004}

In addition, we have to consider the experimental possibilities for the
requirements in Eqs. (\ref{V12}) and (\ref{V2<J}). Generally, $\langle
s_{2}^{z}\rangle $ must be within $[-0.5,0.5]$ (as shown in Fig. \ref{fig:Tspin of dot 2 J>0};
$|\langle s_{2}^{z}\rangle |$ is actually less
than 0.2 in our results). According to the experiment based on a
semiconductor QD,\cite{Chen2004} $J\sim 0.09$meV, so the order of the spin
biases $V_{1}$ required for the manipulation should $>>0.045$ meV, which
should be within reach of the experiments. For example, Zaffalon and van Wees%
\cite{Zaffalon2003} reported that the spin-polarized current injected from
Co electrodes to an Al island induces a splitting $\mu _{\uparrow }-\mu
_{\downarrow }=eIR_{s}/P$, where $I\sim 10-100\mu $A is the injection
current, $P=7\%$ is the spin injection efficiency of the Co/Al tunnel
barrier, and $R_{s}$ is a defined quantity of the dimension of resistance, which
was measured up to 250m$\Omega $ at $4.2$K. Therefore, the spin bias in this
experiment can be as large as $e\times 0.1\mathrm{mA}\times 250\mathrm{m}%
\Omega /0.07\sim 0.35$meV. Although it is not explicitly addressed in publications\cite{Zhu2001,Motsnyi2002,Hanbicki2002,Kato2004sci,Wunderlich2005,Kotissek2007}
to our knowledge, the values of the spin-dependent chemical
potential splitting in semiconductors may be larger than those in
metals, based on the following two arguments: First, the density of
states near the Fermi level is much smaller in semiconductors.
The same imbalance of one spin component compared to that of the other should
occupy a wider range of energy. Second, the measured spin diffusion
lengths in semiconductors are usually $\sim 1-10\mu
$m,\cite{Kotissek2007} much longer than those in metals ($\sim $
hundreds of nanometers),\cite{Valenzuela2007} so the spin accumulation in
semiconductors are more robust in maintaining a considerable spin bias.
For Eq. (\ref{V2<J}), $V_{2}$ should be, of course, as small as
possible, while the only low bound for $V_{2}$ is that it should be
large enough to generate measurable current. The current measured in a
QD experiment can be as small as $\Delta I\sim $ pA (\onlinecite{Vaart1995})
to our knowledge, while the conductance through a quantum dot is on
the order of $\frac{2e^{2}}{h}\approx 7.75\times 10^{-5}\Omega
^{-1}$. Thus, the minimum bias voltage required to generate such a
small current is on the order of $\Delta
I/\frac{2e^{2}}{h}=\mathrm{pA}/(7.75\times 10^{-5}\Omega ^{-1})\sim
\mu $V, which is orders of magnitude smaller than the reported value of $J\sim
90\mu $eV estimated in the experiment.\cite{Chen2004}

Throughout this work, the temperature is assumed to be the smallest
one among all physical quantities. According to the
experiment,\cite{Chen2004} we choose $k_{\mathrm{B}}T=4\mu $eV in
the following calculations.

For a brief summary, we compare our spin sensing scheme with the charge
sensing technique in Table. \ref{tab:summary}.

\begin{table}[h]
\caption{Comparison between charge sensing technique and spin sensing
technique in the present work.}
\label{tab:summary}%
\begin{ruledtabular}
\begin{tabular}{cccccccc}
              & Charge sensing\cite{Sprinzak2002,Elzerman2003} & Our model  \\   \hline\\
             Configuration & QD+QPC &QD1+QD2 \\
             & \includegraphics[width=0.15\textwidth]{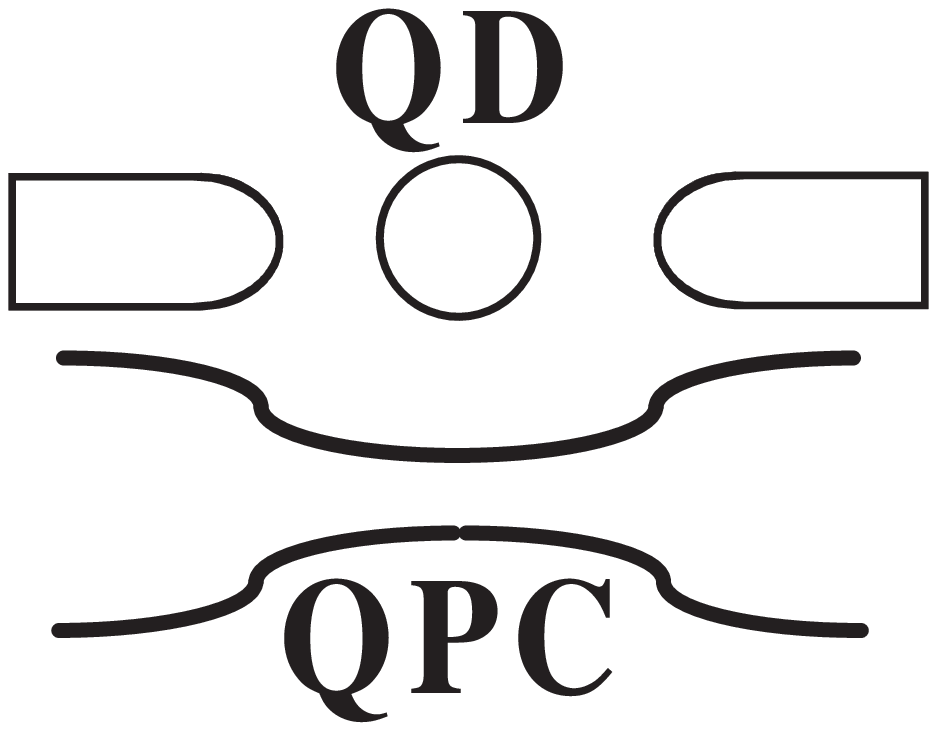} &
             \includegraphics[width=0.15\textwidth]{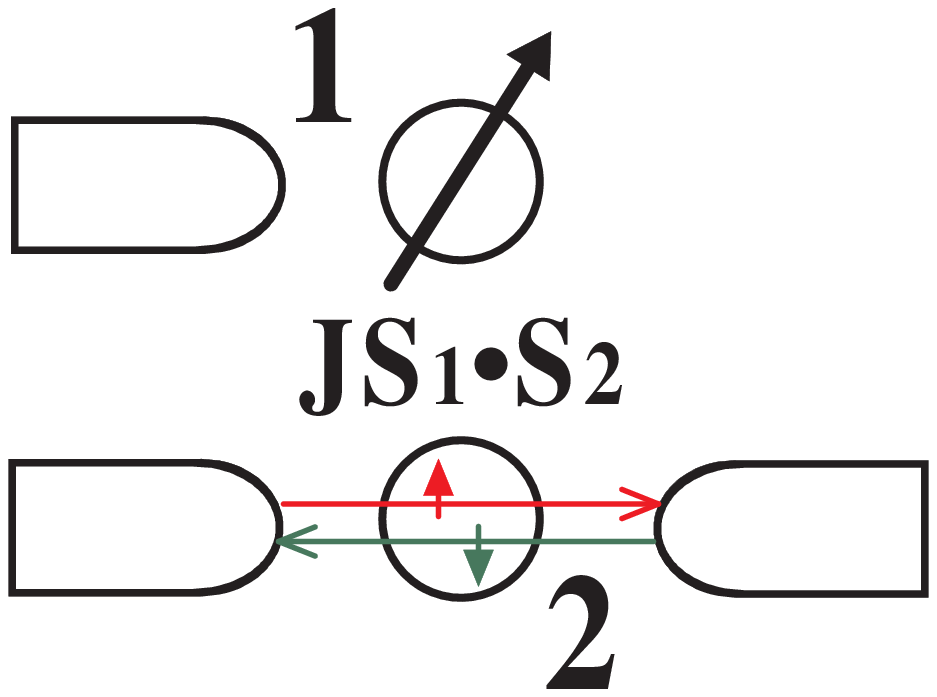}\\\hline
            Quantity &  & \\
            manipulated & Electron number & Single spin\\
            and detected      &  in QD         &     in
            QD1 \\
            (Range)& (0,1,2...) & (-0.5 $\sim $ 0.5)
            \\ \hline
            Manipulation & Tuning gate& Applying large\\ approach     &  voltage of QD        &  spin bias to QD1 \\
            \hline
            Interaction    & Coulomb  & Exchange interaction  \\
             & repulsion & $J\mathbf{s}_1\cdot \mathbf{s}_2 $ \\ \hline
            Detector & QPC &
            QD2 \\ \hline
            Measured & Charge conductance  & Spin conductance\\
            quantity & by applying & by applying  \\
            in detector &  charge bias &  spin bias
 \end{tabular}
\end{ruledtabular}
\end{table}

\section{\label{Sec:dot 1}Manipulation of quantum spin by means of spin bias}

In this section, we present the calculation and numerical results of
the first part of our scheme, the manipulation of the spin in dot 1.
Based on the agreement in the Sec. \ref{sec:validity}, It is safe and
convenient to ignore dot 2 and the spin correlation between dots 1
and 2 in this section.

\subsection{\label{Sec:GF1}Spin polarization and Green's function technique}

In order to determine the optimal parameters to polarize a single spin, we
calculate the polarization of the electron spin $\langle s_{1}^{z}\rangle =
(\langle n_{1\uparrow }\rangle -\langle n_{1\downarrow }\rangle )/2$ and the
total electron number $\langle n_{1}\rangle =\langle n_{1\uparrow }\rangle
+\langle n_{1\downarrow }\rangle $ in dot 1 as functions of the spin
bias $V_{1}$ and the middle point $\mu_{1}$ of the Fermi level. The formula
for the $\sigma $ component of particle number $\langle n_{1\sigma }\rangle $
can be expressed in terms of the lesser Green's function,\cite{Haug1996,Mahan1990}
\begin{equation*}
\langle n_{1\sigma }\rangle =-i\int \frac{d\omega }{2\pi }G_{1\sigma
}^{<}(\omega ),
\end{equation*}
where
\begin{equation*}
G_{1\sigma}^{<}=G_{1\sigma}^{r}\Sigma_{1\sigma}^{<}[{G}_{1\sigma}^{r}]^{\dag
},\ \Sigma_{1\sigma}^{<}=i\Gamma_{1} f_{1\sigma },
\end{equation*}
where $\mathbf{G}^{r}_{1\sigma}(\omega)$ are the retarded Green's functions
defined as the Fourier transform of $G_{1\sigma}^{r}(t)=-i\theta (t)\langle
\{d_{1\sigma}(t),d_{1\sigma }^{\dag }\}\rangle $, where $d_{1\sigma
}(t)=e^{iH_{1}t}d_{1\sigma }e^{-iH_{1}t}$. $\Gamma _{1}=\sum_{k}2\pi
|V_{k\sigma }|^{2}2\pi \delta (\omega -\epsilon _{k\sigma })$ is the
broadening of the quantum dot level $\epsilon _{1}$, due to its coupling to lead L1 for $\uparrow $ or $\downarrow $ electrons. $\Gamma _{1}$ is
assumed to be independent of $\sigma $ since we are not addressing a
ferromagnetic electrode. $f_{1\sigma }$ is the Fermi-Dirac distribution of $\sigma $ electrons in the lead,
\begin{equation*}
f_{1\sigma }(\omega )=\frac{1}{e^{(\omega -\mu _{1\sigma })/k_{B}T}+1},
\end{equation*}%
where $T$ is the temperature and $k_{\mathrm{B}}$ is the Boltzmann
constant. Up to the second-order of Hartree-Fock approximation, the
retarded Green's function of dot 1 is given by\cite{Haug1996}
\begin{equation*}
G_{1\sigma}^{r}=\frac{1-\langle n_{1\overline{\sigma }}\rangle }{\omega
-\epsilon _{1}+\frac{i}{2}\Gamma _{1}}+\frac{\langle n_{1\overline{\sigma }%
}\rangle }{\omega -\epsilon _{1}-U_{1}+\frac{i}{2}\Gamma _{1}}.
\end{equation*}%
Here we do\ not consider the Kondo effect in dot because it is usually
suppressed due to the large spin bias $V_{1}$.

\subsection{Numerical results}

\begin{figure}[h]
\centering \includegraphics[width=0.35\textwidth]{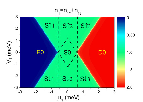} %
\includegraphics[width=0.43\textwidth]{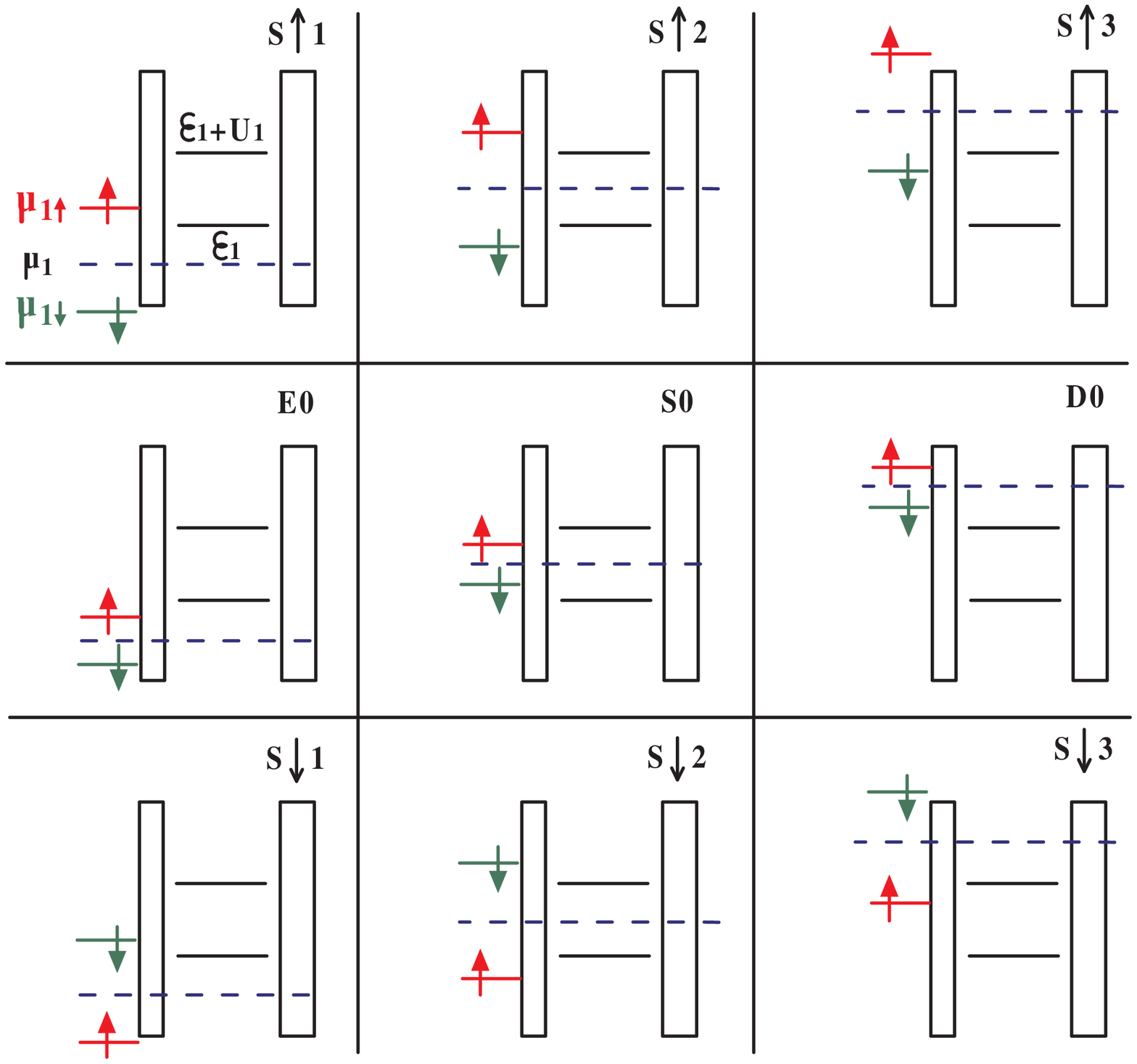} %
\includegraphics[width=0.35\textwidth]{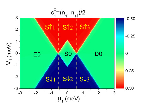}
\caption{The total electron number $\langle
n_{1}\rangle$ (top panel) and spin polarization $\langle
s_{1}^{z}\rangle $ (bottom panel) diagrams as functions of the spin
bias $V_{1}$ and the Fermi level middle point $ \mu_1$. The diagrams
can be divided into nice configurations, schematically shown in the
middle panel, starting from three unpolarized states \textbf{E0}, \textbf{S0}, and \textbf{D0} with
electron numbers $n_1=0,1,2$, respectively. \textbf{E}=empty, \textbf{S}=singly-occupied, \textbf{D}=doubly-occupied; and 0, $\uparrow $, and $\downarrow $ represent unpolarized and two polarized states. }
\label{fig:9 configurations}
\end{figure}

The electron number $\langle n_1\rangle$ and spin-polarization $\langle
s_1^z\rangle$ diagrams are shown in Fig. \ref{fig:9 configurations} as
functions of the pure spin bias $V_1$ and the Fermi level middle point $\mu_1 $. In the calculations, we choose a set of typical parameters $k_{B}T=0.004 $ meV, $U_1=1.2$ meV, and $\Gamma_{1}=0.0375$ meV, in accordance
with the experiment.\cite{Chen2004} $\epsilon_1=-0.6$ meV to assure that $\epsilon_1+U_1/2$ is the energy zero point.

As shown in Fig. \ref{fig:9 configurations}, when the spin bias $V_{1}=0$,
the electron number will be 0, 1, and 2 if $\mu _{1}$ is well below $\epsilon_{1}$,
between $[\epsilon _{1},\epsilon _{1}+U_{1}]$, and above $\epsilon _{1}+U_{1}$, respectively.
The empty, singly occupied, and doubly occupied
unpolarized regimes are denoted by \textbf{E}0, \textbf{S}0, and \textbf{D}
0, respectively. Each regime develops into two spin-polarized regimes when a
positive ($V_{1}>0$) or negative ($V_{1}<0$) spin bias is applied. \textbf{E}
0 $\rightarrow$ \textbf{S}$\uparrow$1, \textbf{S}$\downarrow$1; \textbf{S}0 $
\rightarrow$ \textbf{S}$\uparrow$2, \textbf{S}$\downarrow$2; and \textbf{D}0 $
\rightarrow$ \textbf{S}$\uparrow$3, \textbf{S}$\downarrow$3. We denote the nine
regimes by the electron number and polarization. 0, $\uparrow $, and $
\downarrow $ stand for unpolarized, spin up, and spin down, respectively. For
example, $\mathbf{S}$$\mathbf{\uparrow }$$\mathbf{2}$ represents the second
regime when dot 1 is singly occupied and $\uparrow $ polarized. We
describe the nine regimes one by one as follows:

(1) \textbf{E0}: Both $\mu_{1\uparrow}$ and $\mu_{1\downarrow}$ are well below $
\epsilon_1$; the dot is then empty and unpolarized.

(2) \textbf{S}$\uparrow $\textbf{1}: This is obtained by applying a positive spin bias to \textbf{E0} until $\mu
_{1\uparrow }$ is above $\epsilon _{1}$, while $\mu _{1\downarrow }$ is still
below $\epsilon _{1}$, so only an electron of spin up is energetically
allowed to occupy dot. The spin in dot 1 is $\uparrow $ polarized.

(3) \textbf{S}$\downarrow $\textbf{1}: This is opposite to \textbf{S}$\uparrow $1 and is obtained by applying a
negative spin bias.

(4) \textbf{S0}: Both $\mu _{1\uparrow }$ and $\mu _{1\downarrow }$ are between $%
\epsilon _{1}$ and $\epsilon _{1}+U_{1}$. At least one electron can be
filled into dot. However, neither $\mu _{1\uparrow }$ nor $\mu _{1\downarrow }$
can compensate for the charge energy $U_{1}$ for filling the second
electron. The opportunity of occupation for spin-up or spin-down electron is the
same. Therefore, the electron spin is unpolarized.

(5) \textbf{S}$\uparrow $\textbf{2}: This is obtained by applying a positive spin bias to \textbf{S0} until $\mu
_{1\downarrow }$ is well below $\epsilon _{1}$ or $\mu _{1\uparrow }$ is
well above $\epsilon _{1}+U_{1}$. The former situation is similar to \textbf{%
S}$\uparrow $1. In the latter situation, if dot is initially occupied by
a $\downarrow $ spin, an $\uparrow $ spin can still enter dot because $%
\mu _{1\uparrow }>\epsilon _{1}+U_{1}$. Once the $\uparrow $ spin enters the dot,
the $\downarrow $ spin will be repulsed out of dot and unable to enter again
because $\mu _{1\downarrow }<\epsilon _{1}+U_{1}$ cannot supply enough
charge energy. Both situations lead to a spin $\uparrow $ electron filled in
dot.

(6) \textbf{S}$\downarrow $\textbf{2}: This is opposite to \textbf{S}$\uparrow $2 and is obtained by applying a
negative spin bias.

(7) \textbf{D0}: Both $\mu _{1\uparrow }$ and $\mu _{1\downarrow }$ are well above
$\epsilon _{1}+U_{1}$; dot is occupied by two electrons, one up and the
other down due to the Pauli exclusive principle. Thus, there is no polarization.

(8) \textbf{S}$\uparrow $\textbf{3}: $\mu_{1\uparrow}$ is well above $\epsilon_1+U_1$, while
$\mu_{1\downarrow}$ is below $\epsilon_1+U_1$.

(9) \textbf{S}$\downarrow $\textbf{3}: This is opposite to \textbf{S}$\uparrow $3.

\begin{figure}[h]
\centering \includegraphics[width=0.45\textwidth]{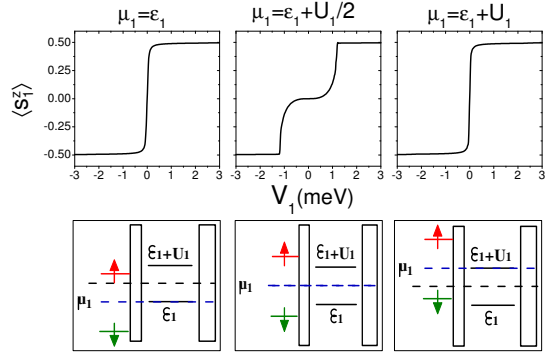}
\caption{$\langle s_{1}^{z}\rangle $ vs $V_{1}$ when $ %
\mu _{1}$ is aligned with $ \epsilon _1$, $ \epsilon_1+U_1/2 $%
, and $ \epsilon_1+U_1$. Notice that for the first and the last
cases, $\langle s_{1}^{z}\rangle $ is immediately reversed between
$\sim 1/2$ and $\sim -1/2$ when spin bias $V_1$ is changed from
positive to negative over a width $\sim \Gamma_1$. }
\label{fig:polarization of dot 1}
\end{figure}

As shown in Fig. \ref{fig:9 configurations}, the edges between different
regimes are very sharp because $\Gamma_1$ is much smaller than $U_1$.
Increasing dot-lead coupling will blur the edges, and lower the
efficiency of manipulation. Moreover, as shown in Fig. \ref{fig:polarization
of dot 1}, at two conditions when $\mu _{1}$ is aligned with $\epsilon _{1}$%
, or $\epsilon _{1}+U_{1}$, the maximal polarization can be achieved
immediately as long as the spin bias $V_{1}$ overwhelms $\Gamma _{1}$, the
broadening of dot level due to dot-lead coupling. In both situations
there is only one electron in dot. So if dot is weakly coupled to
the leads, one can manipulate the single spin by using a small spin bias. $%
\mu _{1}=\epsilon _{1}+U_{1}/2$ in Fig.\ref{fig:polarization of dot 1}
represents the hardest condition to polarize the spin, in which the spin
bias has to overcome the intradot Coulomb repulsion energy.

\section{\label{Sec:detection}Measurement of quantum spin by means of spin
bias}

In this section, the calculation and numerical results on the transport
through dot 2 in the presence of a small spin bias $V_2$ is
presented, and its relation with the spin orientation in dot 1 is
illustrated in details.

\subsection{\label{Sec:effectivefield}Effective field approximation}

Based on Eqs. (\ref{V12}) and (\ref{V2<J}), when calculating the physical
quantities of dot 2, the self-consistency for those of dot 1 can be
ignored. In this limit, we treat the spin of dot 1 as an effective
magnetic field using the Hamiltonian
\begin{equation}
V_{12}^{\prime}=\frac{J}{2}s_{1}^{-}d_{2\uparrow }^{\dag }d_{2\downarrow }+%
\frac{J}{2}s_{1}^{+}d_{2\downarrow }^{\dag }d_{2\uparrow }+\frac{J}{2}%
s_{1}^{z}(n_{2\uparrow }-n_{2\downarrow }),
\end{equation}%
where $s_{1}^{z,+,-}$ are quantum field operators. It is worth stressing
that the self-consistent calculation is very complicated, but still
available in the present problem. To avoid mathematics, the present
approximation is believed to give an intuitive picture of how the spin bias is
converted into a charge signal.

\subsection{\label{Sec:GF2}Spin-dependent transport and Green's function
technique}

In Sec. \ref{sec:scheme}, we have made it clear that the direction of the net
charge current that can flow through dot 2 in the presence of a small
spin bias $V_{2}$ can be used to detect the spin orientation in dot 1.
Thus, it is convenient to introduce a quantity spin conductance, which is
defined as the ratio of the charge current to the spin bias in the limit of
the zero spin bias.\cite{Wang2004} In Appendix \ref{Sec:spinconductance}, we
derive the formula of spin conductance at zero temperature,
\begin{equation}
\mathcal{G}^{s}(\mu )=\lim_{V^{s}\rightarrow 0}\frac{\partial (I_{\uparrow
}+I_{\downarrow })}{\partial V^{s}}=\frac{e^{2}}{h}[T_{\uparrow }(\mu
)-T_{\downarrow }(\mu )],
\end{equation}%
where $T_{2\sigma }$ is the transmission probability for $\sigma $ electrons
through dot 2 within the framework of linear response theory. Therefore, the
numerical results presented in Secs. \ref{Sec:dot 2} and \ref{Sec:J<0} are
given only in terms of $T_{2\uparrow }-T_{2\downarrow }$ in the unit of $%
\frac{e^{2}}{h}$.

In the linear response theory, the $\sigma $ component of the transmission
probability is given by\cite{Datta1997}
\begin{equation*}
T_{2\sigma }(\mu _{2})=[G_{2\sigma }^{r}(\omega )\Gamma _{2}^{L}(G_{2\sigma
}^{r}(\omega ))^{\dag }\Gamma _{2}^{R}]_{\omega =\mu _{2}},
\end{equation*}%
where the retarded Green's function $G_{2}^{r}(\omega )$ is defined as the
Fourier transform of
\begin{equation*}
G_{2\sigma }^{r}(t)=-i\theta (t)\langle \{d_{2\sigma }(t),d_{2\sigma }^{\dag
}\}\rangle
\end{equation*}
with $d_{2\sigma }(t)=e^{i(H_{2}+V_{12})t}d_{2\sigma }e^{-i(H_{2}+V_{12})t}$%
. $\Gamma _{2}^{\alpha }=\sum_{k}2\pi |V_{k\alpha \sigma }|^{2}2\pi \delta
(\omega -\epsilon _{k\alpha \sigma })$ is the broadening of dot 2 level $%
\epsilon _{2}$ due to the couplings to the $\alpha $ lead. Using the
equation of motion of $G_{2\sigma }^{r}$ with respect to $%
H_{2}+V_{12}^{\prime}$, we arrive at (detailed deductions and approximations
are given in Appendix \ref{Sec:deduction of G2})%
\begin{equation*}
\left(
\begin{array}{ccc}
\Omega & -\frac{J}{2} & -\frac{J}{2} \\
-\frac{J}{8} & \Omega & \frac{J}{4} \\
-\frac{J}{4} & \frac{J}{2} & \Omega +\frac{J}{4}%
\end{array}%
\right) \left(
\begin{array}{c}
G_{2\uparrow }^{r} \\
G_{2} \\
G_{3}%
\end{array}%
\right) =\left(
\begin{array}{c}
1+UG_{4} \\
\langle s_{1}^{z}\rangle +UG_{5}+\frac{J}{2}G_{6} \\
JG_{5}+\frac{J+2U}{2}G_{6}%
\end{array}%
\right) ,
\end{equation*}

\begin{equation*}
\left(
\begin{array}{ccc}
\Omega -U & -\frac{J}{2} & -\frac{J}{2} \\
-\frac{J}{8} & \Omega -U & -\frac{J}{4} \\
-\frac{J}{4} & -\frac{J}{2} & \Omega -\frac{J+4U}{4}%
\end{array}%
\right) \left(
\begin{array}{c}
G_{4} \\
G_{5} \\
G_{6}%
\end{array}%
\right) =\left(
\begin{array}{c}
N_{1} \\
N_{2} \\
N_{3}%
\end{array}%
\right) ,
\end{equation*}

\begin{equation*}
\left(
\begin{array}{ccc}
\Omega & \frac{J}{2} & -\frac{J}{2} \\
\frac{J}{8} & \Omega & -\frac{J}{4} \\
-\frac{J}{4} & -\frac{J}{2} & \Omega +\frac{J}{4}%
\end{array}%
\right) \left(
\begin{array}{c}
G_{2\downarrow }^{r} \\
G_{8} \\
G_{9}%
\end{array}%
\right) =\left(
\begin{array}{c}
1+UG_{10} \\
\langle s_{1}^{z}\rangle +UG_{11}-\frac{J}{2}G_{12} \\
-JG_{11}+\frac{J+2U}{2}G_{12}%
\end{array}%
\right) ,
\end{equation*}

\begin{equation}
\left(
\begin{array}{ccc}
\Omega -U & \frac{J}{2} & -\frac{J}{2} \\
\frac{J}{8} & \Omega -U & \frac{J}{4} \\
-\frac{J}{4} & \frac{J}{2} & \Omega -\frac{J+4U}{4}%
\end{array}%
\right) \left(
\begin{array}{c}
G_{10} \\
G_{11} \\
G_{12}%
\end{array}%
\right) =\left(
\begin{array}{c}
N_{4} \\
N_{5} \\
N_{6}%
\end{array}%
\right) ,  \label{G2}
\end{equation}%
where we denote $\Omega =\omega -\epsilon _{2}+\frac{i}{2}\Gamma _{2}$, and $
\Gamma _{2}=\Gamma _{2}^{L}+\Gamma _{2}^{R}$. $G_{2}=\langle \langle
d_{2\uparrow }s_{1}^{z}|d_{2\uparrow }^{\dag }\rangle \rangle $, $%
G_{3}=\langle \langle d_{2\downarrow }s_{1}^{-}|d_{2\uparrow }^{\dag
}\rangle \rangle $, $G_{4}=\langle \langle d_{2\uparrow }n_{2\downarrow
}|d_{2\uparrow }^{\dag }\rangle \rangle $, $G_{5}=\langle \langle
d_{2\uparrow }n_{2\downarrow }s_{1}^{z}|d_{2\uparrow }^{\dag }\rangle
\rangle $, $G_{6}=\langle \langle d_{2\downarrow }n_{2\uparrow
}s_{1}^{-}|d_{2\uparrow }^{\dag }\rangle \rangle $, $G_{8}=\langle \langle
d_{2\downarrow }s_{1}^{z}|d_{2\downarrow }^{\dag }\rangle \rangle $ , $%
G_{9}=\langle \langle d_{2\uparrow }s_{1}^{+}|d_{2\downarrow }^{\dag
}\rangle \rangle $, $G_{10}=\langle \langle d_{2\downarrow }n_{2\uparrow
}|d_{2\downarrow }^{\dag }\rangle \rangle $, $G_{11}=\langle \langle
d_{2\downarrow }n_{2\uparrow }s_{1}^{z}|d_{2\downarrow }^{\dag }\rangle
\rangle $, and $G_{12}=\langle \langle d_{2\uparrow }n_{2\downarrow
}s_{1}^{+}|d_{2\downarrow }^{\dag }\rangle \rangle $, where the notation $%
\langle \langle A|B\rangle \rangle $ stands for the Fourier transform of the
retarded Green's function $-i\theta (t)\langle \{A(t),B\}\rangle $. $%
N_{1}=\langle n_{2\downarrow }\rangle $, $N_{2}=\langle n_{2\downarrow
}s_{1}^{z}\rangle $, $N_{3}=-\langle d_{2\uparrow }^{\dag }d_{2\downarrow
}s_{1}^{-}\rangle $, $N_{4}=\langle n_{2\uparrow }\rangle $, $N_{5}=\langle
n_{2\uparrow }s_{1}^{z}\rangle $, and $N_{6}=-\langle d_{2\downarrow }^{\dag
}d_{2\uparrow }s_{1}^{+}\rangle $. It is worth pointing out that our
deduction is similar to that in a previous work,\cite{Tolea2007} however, our
calculation retains $\langle s_{1}^{z}\rangle $ as an input parameter
ranging from -0.5 to 0.5, reflecting the spin polarization in dot 1 to
be detected.

Using the identity of Green's functions at equilibrium,\cite{Mahan1990,Haug1996}
\begin{equation*}
\mathbf{G}^{<}=i\mathbf{A}f,\ \ \mathbf{A}=-2\mathrm{Im}\mathbf{G}^{r},
\end{equation*}
where $\mathbf{A}$ is the spectral function. We self-consistently calculate the
expectation values in Eq. (\ref{G2}), for example,
\begin{eqnarray}
\langle d_{2\uparrow }^{\dag }d_{2\downarrow }s_{1}^{-}\rangle &=&-i\langle
\langle d_{2\downarrow }(t)s_{1}^{-}(t)|d_{2\uparrow }^{\dag }(t^{\prime
})\rangle \rangle _{t=t^{\prime }}^{<}  \notag  \label{selfconsistence} \\
&=&-i\int \frac{d\omega }{2\pi }\langle \langle d_{2\downarrow
}s_{1}^{-}|d_{2\uparrow }^{\dag }\rangle \rangle _{\omega }^{<}  \notag \\
&= &-\frac{1}{\pi }\int d\omega f_{2}(\omega )\mathrm{Im}\langle \langle
d_{2\downarrow }s_{1}^{-}|d_{2\uparrow }^{\dag }\rangle \rangle ,
\end{eqnarray}%
where
\begin{equation*}
f_{2}=\frac{1}{e^{(\omega -\mu _{2})/k_{B}T}+1}.
\end{equation*}

\subsection{\label{Sec:dot 2}Numerical results}

For numerical calculations, we choose a set of parameters consistent
with the parallel-coupled double-quantum-dot experiment by Chen \emph{et al}%
.,\cite{Chen2004} $k_{\mathrm{B}}T=0.004$ meV, $J=0.09$ meV, $U_{2}=0.8$
meV, and $\Gamma _{2}^{\alpha }=0.0375$ meV. We set $\epsilon _{2}=-0.4$ meV
to assure that $\epsilon _{1}+U_{1}/2=\epsilon _{2}+U_{2}/2$ is the zero
point of energy. Despite the fact that the above experiment parameters lead to the Kondo
effect at low temperatures,\cite{Chen2004} we have enough reasons to rule it
out from the present considerations. For dot 1, the spin is polarized by
the large spin bias $V_{1}$; thus, Kondo effect is usually quenched. For the
dot 2, later we will see that the results are meaningful only when $\mu _{2}$ is
around the $\epsilon _{2}$ and $\epsilon _{2}+U_{2}$, where the first-order
tunneling between leads L2 and R2 through dot 2 is dominant and
suppresses the Kondo effect.

We assume that there is an unknown spin in dot 1 to be detected, i.e.,
all the \textbf{S} regimes in Fig. \ref{fig:9 configurations}. The states of
dot 2 and the double-dot are classified by the electron occupation, and
are shown with their energies in Table \ref{tab:dot2energy} for $J>0$. The
one-electron state of the double-dot is also the empty state (0) of dot
2. The doubly occupied states (D) of dot 2 are two-fold degenerate
triply occupied states of the double-dot. The singly occupied state of the
dot 2, due to the exchange interaction, could favor parallel (P) or
antiparallel (AP) alignment with the spin in dot 1.

\begin{table}[htbp]
\caption{The energy
spectrum of dot 2 for $J>0$ when assuming there is an electron in dot 1,
where P or AP corresponds, respectively, to the situation where parallel or antiparallel alignment of two spins is favored.
One just exchanges the
antiparallel and parallel alignments to have the $J<0$ case. }
\label{tab:dot2energy}%
\begin{ruledtabular}
\begin{tabular}{cccccccc}
electron number\\
 in double-dot & 1 & 2 & 3 \\
electron number\\
    in dot 2  & 0 & 1 & 2   \\\hline
       Ground state     & Empty(0)        &     Anti-parallel(AP)   & Double(D)   \\
            & $E_0=0$        &  $E_1^{\mathrm{AP}}=\epsilon_2-\frac{3}{4}J$   & $E_2=2\epsilon_2+U_2$   \\
       Excited states &  & Parallel(P) &  \\
        &  & $E_1^{\mathrm{P}}=\epsilon_2+\frac{1}{4}J$
 \end{tabular}
\end{ruledtabular}
\end{table}

The poles of Green's functions of dot 2 reflect the energies required
for transitions between these states with different occupancies of
electrons. When the Fermi surface of leads are aligned with these poles and
supply the required energy, the transitions will take place, giving rise to
a conductance peak. We list the four possible transitions to fill dot 2
with $0\rightarrow 1\rightarrow 2$ electrons when $J>0$ in Table \ref{tab:dot2transition},
together with the energies required by the transitions
and the corresponding charge conductance peaks in Fig. \ref{fig:Tcharge of
dot 2 J>0}.

\begin{table}[htbp]
\caption{Transitions between states with different particle numbers in the
dot 2 when $J>0$, energies required from the Fermi surface to supply the
transitions, and the corresponding conductance peaks in Fig.  \ref%
{fig:Tcharge of dot 2 J>0}.}
\label{tab:dot2transition}%
\begin{ruledtabular}
\begin{tabular}{cccccccc}
 &Transitions  & Required energies & Peaks in Fig. \ref{fig:Tcharge
of dot 2 J>0}
\\\hline
&0 $\rightarrow$ AP & $E_1^{\mathrm{AP}}-E_0=\epsilon_2-\frac{3}{4}J$ &  1 (left)\\
&0 $\rightarrow$ P     & $E_1^{\mathrm{P}}-E_0=\epsilon_2+\frac{1}{4}J$ & 2 \\
& P $\rightarrow$ D     & $E_2-E_1^{\mathrm{P}}=\epsilon_2+U_2-\frac{1}{4}J$ & 3\\
& AP $\rightarrow$ D &  $E_2-E_1^{\mathrm{AP}}=
\epsilon_2+U_2+\frac{3}{4}J$ & 4
(right)\\
\end{tabular}
\end{ruledtabular}
\end{table}

\begin{figure}[htbp]
\centering \includegraphics[width=0.4\textwidth]{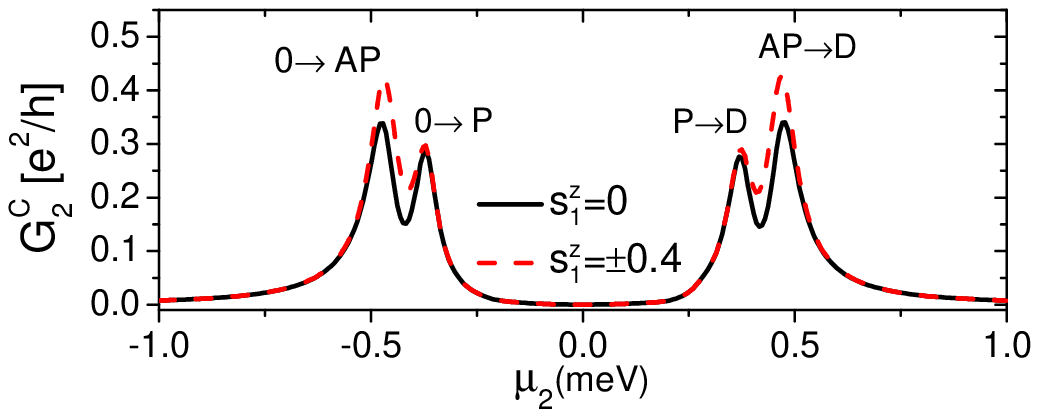}\newline
\includegraphics[width=0.45\textwidth]{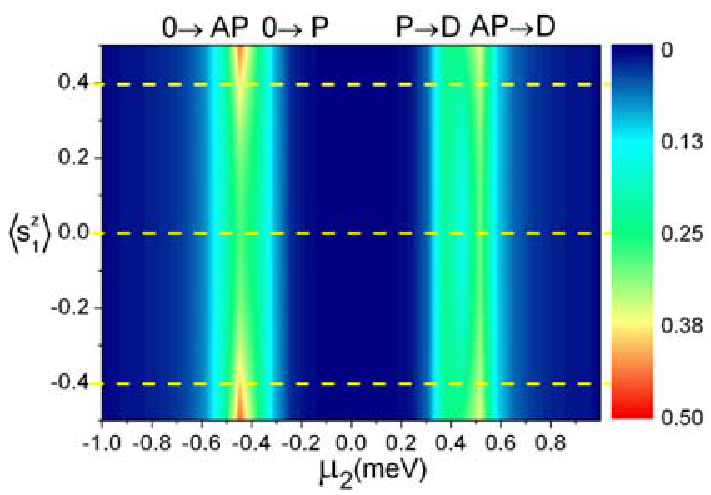}\newline
\caption{The charge conductance ( $\frac{e^2}{h}(T_{2\uparrow
}+T_{2\downarrow})$ ) vs $ \mu_{2}$ and $\langle s_{1}^{z}\rangle $. $%
 \mu _{2}$ is the equilibrium Fermi level. The exchange coupling
strength $J=0.09$ meV, $U_{2}=0.8$ meV, $ \epsilon _{2}=-0.4$ meV,
and $\Gamma _{2}^{ \alpha}=0.0375$ meV. The upper panel shows the values
along the horizontal dashed lines in the lower panel. }
\label{fig:Tcharge of dot 2 J>0}
\end{figure}

\begin{figure}[htbp]
\centering \includegraphics[width=0.45\textwidth]{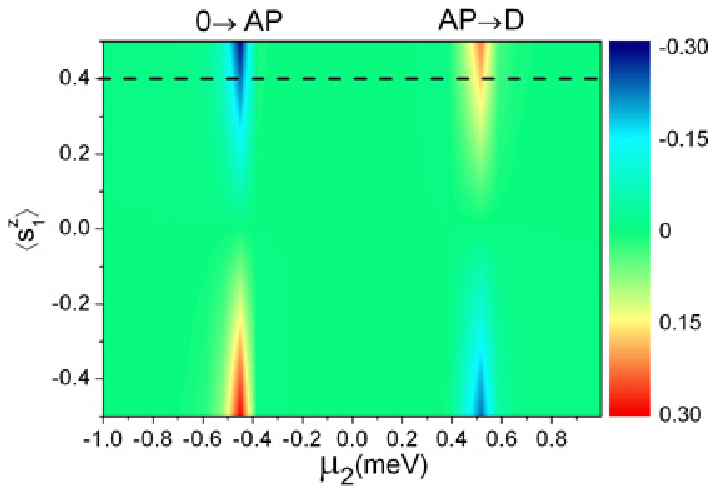}\newline
\includegraphics[width=0.4\textwidth]{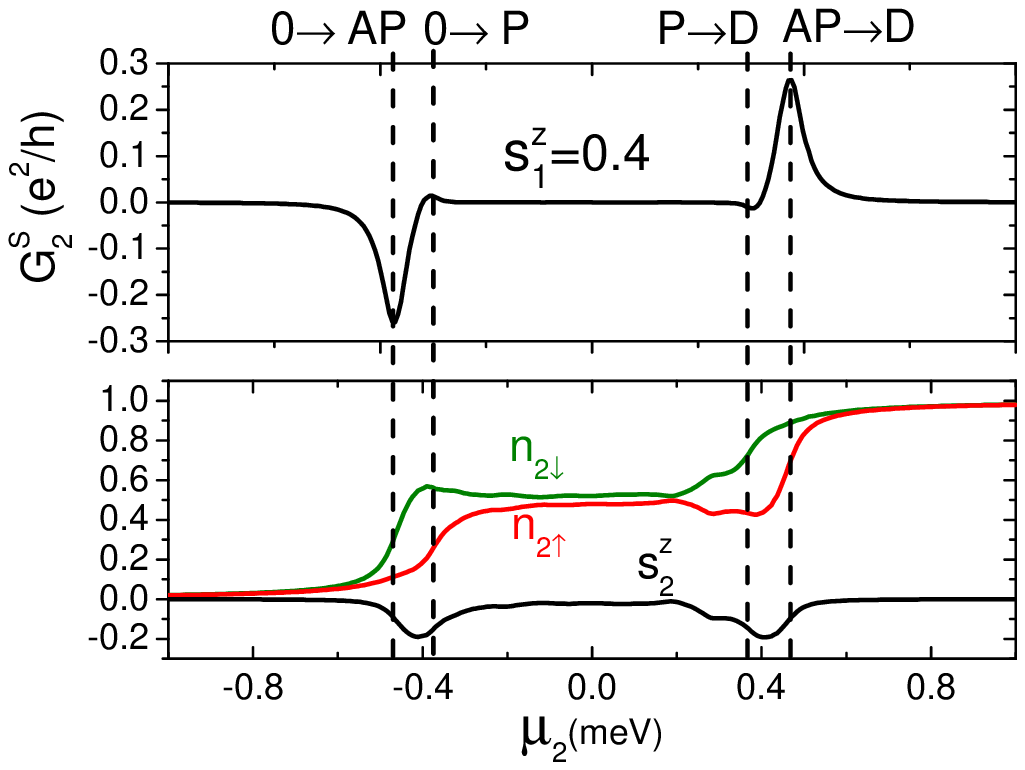}\newline
\caption{The spin conductance (
$\frac{e^2}{h}(T_{2\uparrow }-T_{2\downarrow })$ ) vs $\mu_{2}$ and
$\langle s_{1}^{z}\rangle $. $\mu_{2}$ is the equilibrium Fermi
level. The parameters are the same as those in Fig. \ref{fig:Tcharge
of dot 2 J>0}. The lower panel shows the values along the horizontal
dashed line in the higher panel. } \label{fig:Tspin of dot 2 J>0}
\end{figure}

There are four charge conductance peaks in Fig. \ref{fig:Tcharge of dot 2
J>0}, forming two groups spaced by $U_{2}$. The peaks in each group are
separated by $J$. When $\langle s_{1}^{z}\rangle =0$, the numerical results
are in good agreement with those by Tolea and Bulka. \cite{Tolea2007} From
the above results, one understands why we ignore the inter-dot capacitive
repulsion $U^{\prime }n_{1}n_{2}$. In the singly occupied regime of dot
1, this term adds $U^{\prime }$ to the singly occupied energy and $2U^{\prime }$
to the doubly occupied energy of dot 2, so it only widens
the spacing between peaks 1 and 2 with respect to peaks 3 and 4 by $U^{\prime }$ in
the conductance spectrum of dot 2 and does not contribute to any
spin-dependent effect.

Interesting results emerge when $\langle s_{1}^{z}\rangle \neq 0$. As an
example, we consider the case $\langle s_{1}^{z}\rangle >0$; i.e., an $\uparrow $
electron is in dot 1. We start with the empty state of the
dot 2; i.e., $\mu _{2}$ is well below the energy of $0\rightarrow $ AP
transition at $\epsilon _{2}-3J/4$. When $\mu _{2}$ is raised to be aligned
with the transition pole $0\rightarrow $ AP, dot 2 will favor $%
\downarrow $ electron occupation because of the $\uparrow$ electron in the
dot 1. As shown in Fig. \ref{fig:Tspin of dot 2 J>0}, the difference $%
(n_{2\downarrow }-n_{2\uparrow })$ and $\langle s_{2}^{z}\rangle $ reach
maximum after $\mu _{2}$ is above the transition $0\rightarrow $ AP at $%
\epsilon _{2}-3J/4$. The transport of $\downarrow $ ($\uparrow $) electrons
through dot 2 via $0\rightarrow$ AP thus will be enhanced (suppressed),
i.e., $T_{2\uparrow }-T_{2\downarrow }<0$, which accounts for the negative
value region around $\epsilon _{2}-3J/4$ (transition 0 $\rightarrow $ AP) in
Fig. \ref{fig:Tspin of dot 2 J>0} when $\langle s_{1}^{z}\rangle >0$.
When raising of $\mu _{2}$ is continued until it is aligned with the transition $0\rightarrow $
P at $\epsilon _{2}+J/4$, the electron in dot 2 could be either $%
\uparrow $ or $\downarrow $ because both situations are energetically
allowed. A direct result of this nearly arbitrary spin polarization is that
the second electron added to dot 2 via the transition P $\rightarrow $ D
can also be either $\uparrow $ or $\downarrow $. Thus, no spin orientation is
particularly favored when electrons tunnel via the transitions 0 $%
\rightarrow $ P at $\epsilon _{2}+\frac{1}{4}J$ and P $\rightarrow $ D at $%
\epsilon _{2}+U_{2}-\frac{1}{4}J$. As a result, $n_{2\downarrow
}-n_{2\uparrow }$ approaches zero between $\epsilon _{2}+\frac{1}{4}J$
and $\epsilon _{2}+U_{2}-\frac{1}{4}J$, and there is only an invisible
difference between $T_{2\uparrow }$ and $T_{2\downarrow }$ at both energies.
If the second electron is added via the transition AP $\rightarrow $ D at $%
\epsilon _{2}+U_{2}+\frac{3}{4}J$, it will automatically favor $\uparrow $
because there is already a $\downarrow $ electron in dot 2. This process
is clearly shown as $n_{2\downarrow }-n_{2\uparrow }$ reaches a maximum
between $\epsilon _{2}+U_{2}-\frac{1}{4}J < \mu _{2} < \epsilon _{2}+U_{2}+%
\frac{3}{4}J$ and finally goes to zero after $\mu _{2}>\epsilon _{2}+U_{2}+%
\frac{3}{4}J$. Hence, $T_{2\uparrow }-T_{2\downarrow }>0$ when $\mu _{2}$ is
aligned with the transition AP $\rightarrow $ D at $\epsilon _{2}+U_{2}+%
\frac{3}{4}J$.

\begin{figure}[htbp]
\centering \includegraphics[width=0.42\textwidth]{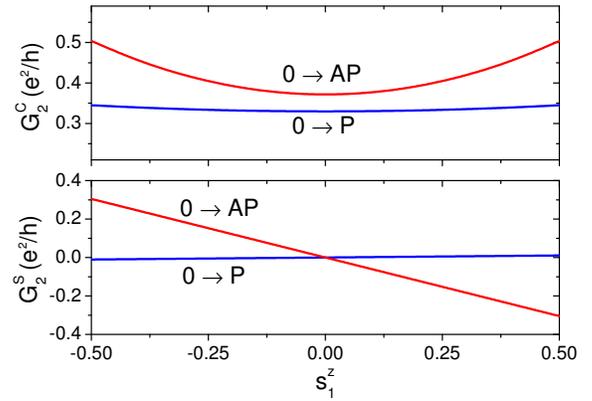}
\caption{The charge conductance ($T_{2\uparrow
}+T_{2\downarrow }$) and spin conductance ($T_{2\uparrow }-T_{2\downarrow }$%
) vs $\langle s_{1}^{z}\rangle $ when $\mu _{2}= \epsilon %
_{2}-3J/4$($0\rightarrow $AP) and $\epsilon _{2}+J/4$ ($0\rightarrow
$ P). The other parameters are the same as in Figs. \ref{fig:Tcharge
of dot 2 J>0} and \ref{fig:Tspin of dot 2 J>0}.} \label{fig:T2 vs
s1z}
\end{figure}

By the same token, the case $\langle s_{1}^{z}\rangle <0$ can be calculated.
In Fig. \ref{fig:T2 vs s1z}, we compare the charge and spin conductances as
functions of $\langle s_{1}^{z}\rangle $ at $\mu _{2}=\epsilon _{2}-3J/4$ ($%
0\rightarrow $ AP) and $\epsilon _{2}+J/4$ ($0\rightarrow $ P). The major
difference is that at $0\rightarrow $ AP, $T_{2\uparrow }-T_{2\downarrow }$
changes sign as $\langle s_{1}^{z}\rangle $ turns from positive to negative
polarization, while $T_{2\uparrow }+T_{2\downarrow }$ remains positive.
Therefore, the spin conductance of dot 2 provides a practical tool to
probe the spin polarization in dot 1.

\subsection{\label{Sec:J<0}Model study when $J<0$}

\begin{figure}[h]
\centering \includegraphics[width=0.35\textwidth]{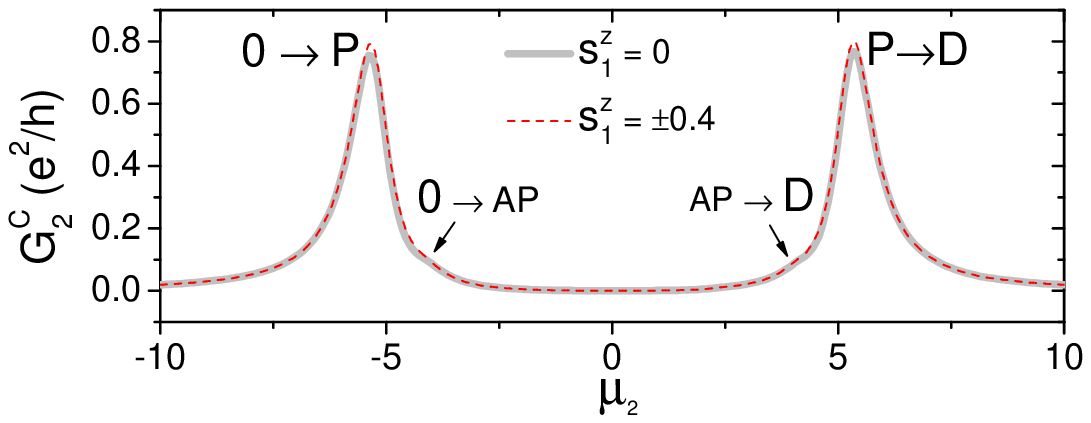} %
\includegraphics[width=0.35\textwidth]{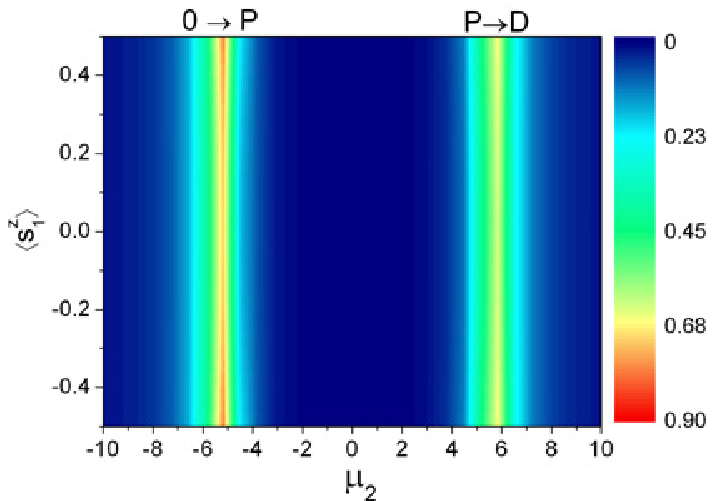} %
\includegraphics[width=0.35\textwidth]{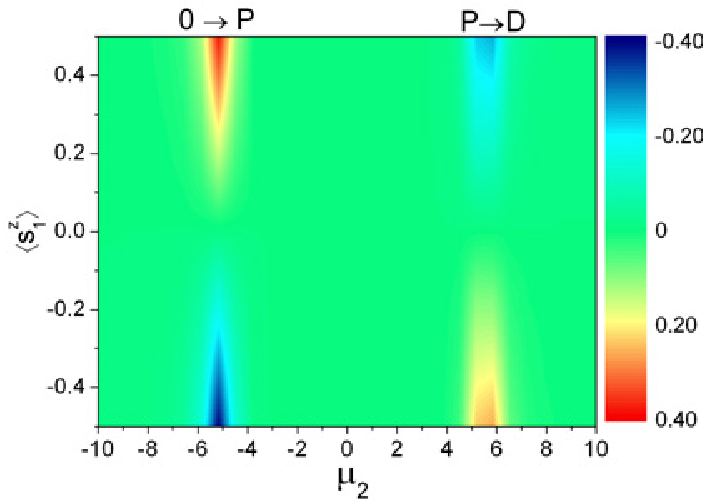} %
\includegraphics[width=0.35\textwidth]{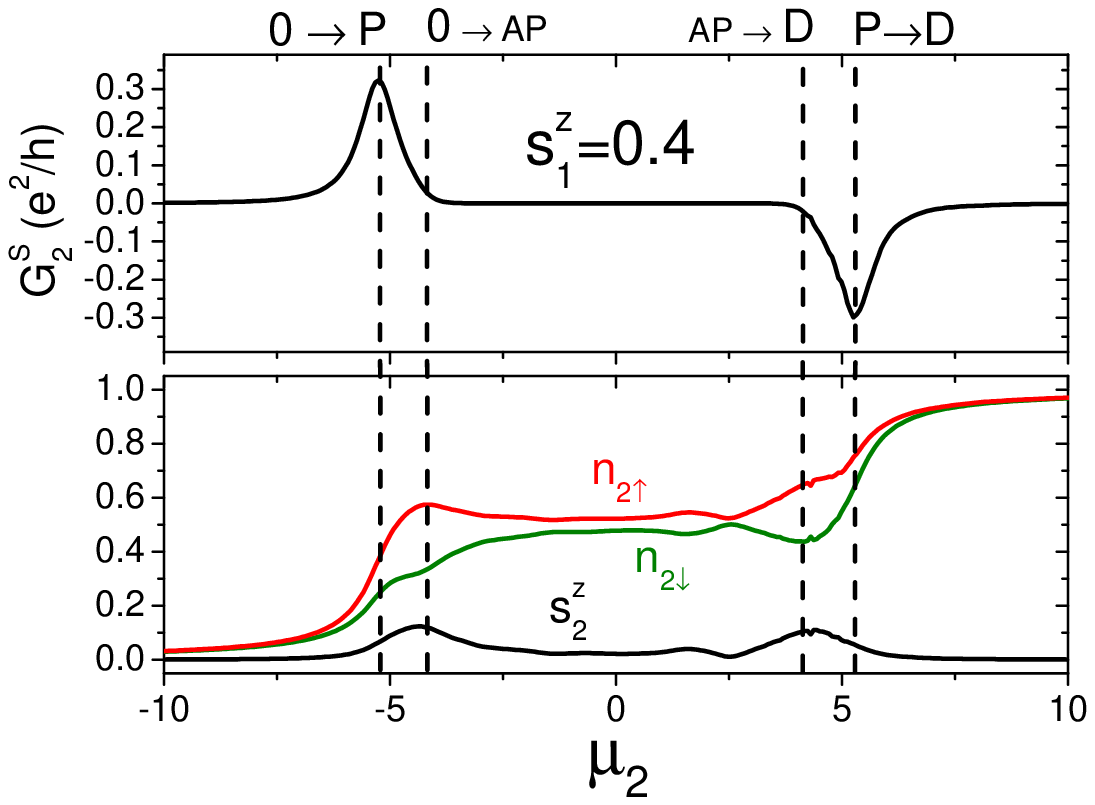}
\caption{The charge conductance [ $\frac{e^2}{h}(T_{2\uparrow
}+T_{2\downarrow })$, top two panels ] and the spin conductance [ $\frac{e^2}{%
h}(T_{2\uparrow }-T_{2\downarrow})$, bottom two panels ] vs $\mu_{2}
$ and $\langle s_{1}^{z}\rangle $ for negative exchange coupling
$J$. The parameters are $J=-1$, $\epsilon_2=-5$, $U_2=10$, and
$\Gamma_2=1$.} \label{fig:T of dot 2 J<0}
\end{figure}

The case of $J<0$ is different from that of $J>0$ because in this situation
the ground states favors the parallel alignment of two spins in the two
dots. Since there are no corresponding experiment data for $J<0$, we just
assume a set of model parameters in analogy to those when $J>0$. In the
charge conductance $\mathcal{G}_{2}^{C}$ in Fig. \ref{fig:T of dot 2 J<0},
only two peaks are clearly visible; they come from the transitions $0\rightarrow $
P and P $\rightarrow $ D. The conductance peaks for $%
0\rightarrow $ AP and AP $\rightarrow $ D are suppressed. Unlike the P
states, which actually originate from the three-fold triplets when $s_{1}^{z}=0$,
there is only one AP state. The contribution of AP to conductance as an
excited state is too weak compared to those of P states, in particular, as
dot 2 is weakly coupled to the leads.\cite{Tolea2007} As we see in Fig. %
\ref{fig:Tcharge of dot 2 J>0}, the peak maximum of the AP state as the
ground state and the total three P states as the excited states are roughly
of the same order, so the peak maximum of 0 $\rightarrow $ AP in $\mathcal{G}%
_{2}^{C}$ of Fig. \ref{fig:T of dot 2 J<0} as the excited state should be
about 1 order ($3\times 3$) smaller than that of the P states as the
ground state.

For the spin conductance $\mathcal{G}_2^S$ in Fig. \ref{fig:T of dot 2 J<0},
two changes occur compared with that in Fig. \ref{fig:Tspin of dot 2 J>0}. The
first is that the peak and dip positions move to $E_{1}^{\mathrm{P}%
}-E_{0}=\epsilon _{2}+\frac{1}{4}J$ and $E_{2}-E_{1}^{\mathrm{P}}=\epsilon
_{2}+U_{2}-\frac{1}{4}J$, because the P states are one-electron ground
states for dot 2 when $J<0$. The second is that the spin conductance
changes sign with respect to that in Fig. \ref{fig:Tspin of dot 2 J>0} because the first
electron that enters dot 2 tends to be parallel aligned with the spin in
dot 1 when $J<0$, in contrast to the anti-parallel alignment when $J>0$.

\section{\label{Sec:summary}Summary}

We proposed a scheme to realize the control and detection of quantum spin in
semiconductor quantum dot by means of spin bias or spin current. A
double-quantum-dot system, coupled to electrodes with spin-dependent
splitting of chemical potentials (spin bias), was investigated to
demonstrate the availability of the proposal. Using a large spin bias, the quantum
spin in dot 1 can be manipulated and maintained in a pure electric
manner. The parameters and regimes of the manipulation were discussed in
details. When an interdot exchange coupling is taken into account, the
ground state of the two spins singly occupied in each dot tends to form an
anti-parallel or a parallel alignment, depending on whether the coupling constant $J$
positive or negative. The spin-dependent transport through dot 2 thus
can be used to detect the polarization of spin in dot 1
nondestructively. We found that the measurement of the spin-dependent
transport can be realized by measuring the net electric current under a spin
bias, which defines a spin conductance. We observed that the spin
conductance of dot 2 changes its sign as the orientation of spin in the
dot 1 reverses, much more sensitively than the usual charge conductance does. The
two cases demonstrate that the spin bias may be a promising approach to
manipulate a single spin while allowing this manipulation to be monitored in
quantum dot systems.

\section{Acknowledgements}

We thank Q. F. Sun, R. L\"{u}, Y. J. Bao, B. Zhou, and R. B. Liu for
discussions. This work was supported by the Research Grant Council of Hong
Kong under Grant No. HKU 7041/07P.

\appendix

\section{The formula of spin conductance}

\label{Sec:spinconductance} The conventional zero-bias differential
conductance for spin component $\sigma $ is defined as
\begin{equation*}
\mathcal{G}_{\sigma }^{c}=\lim_{V^{c}\rightarrow 0}\frac{\partial I_{\sigma }%
}{\partial V^{c}}=-\frac{e}{h}\lim_{V^c\rightarrow 0}\frac{\partial }{%
\partial V^{c}}\int d\omega (f_{L}-f_{R})T_{\sigma }(\omega ),
\end{equation*}%
where without loss of generality the charge bias $V^{c}$ is assumed to
change only the Fermi level of the left lead,
\begin{equation}
f_{L}=\frac{1}{e^{(\omega -\mu +eV^{c})/k_{B}T}+1},f_{R}=\frac{1}{e^{(\omega
-\mu )/k_{B}T}+1},
\end{equation}%
and $\mu $ is the Fermi level of both leads when there is no bias.
Supposing the transmission probability $T_{\sigma }$ is not a function of
the bias,
\begin{equation*}
\mathcal{G}_{\sigma }^{c}=-\frac{e}{h}\int d\omega \lbrack \frac{\partial
f_{L}}{\partial V^{c}}]_{V\rightarrow 0}T_{\sigma }(\omega ),
\end{equation*}%
where $[\frac{\partial f_{L}}{\partial V^{c}}]_{V^{c}\rightarrow
0}\rightarrow -e\delta (\omega -\mu )$ when the temperature $%
k_{B}T\rightarrow 0$. Thus, at zero temperature, the total conductance
including two spin components is
\begin{equation*}
\mathcal{G}^{c}=\lim_{V^{c}\rightarrow 0}\frac{\partial (I_{\uparrow
}+I_{\downarrow })}{\partial V^{c}}=\sum_{\sigma }\mathcal{G}_{\sigma }^{c}=%
\frac{e^{2}}{h}[T_{\uparrow }(\mu )+T_{\downarrow }(\mu )].
\end{equation*}

If a spin bias $V^{s}$ is applied so that $\mu _{\uparrow }^{L}=\mu
_{\downarrow }^{R}=\mu -eV^{s}/2$ and $\mu _{\downarrow }^{L}=\mu _{\uparrow
}^{R}=\mu +eV^{s}/2$, and at zero temperature
\begin{eqnarray}
\lbrack \frac{\partial f_{L\uparrow }}{\partial V^{s}}]_{V^{s}\rightarrow
0}=[\frac{\partial f_{R\downarrow }}{\partial V^{s}}]_{V^{s}\rightarrow 0}
&\rightarrow &-\frac{e}{2}\delta (\omega -\mu ),  \notag \\
\ [\frac{\partial f_{L\downarrow }}{\partial V^{s}}]_{V^{s}\rightarrow 0}=[%
\frac{\partial f_{R\uparrow }}{\partial V^{s}}]_{V^{s}\rightarrow 0}
&\rightarrow &\frac{e}{2}\delta (\omega -\mu ),
\end{eqnarray}
then the differential conductance induced by the spin bias (or spin
conductance for short) is defined as\cite{Wang2004}
\begin{equation}
\mathcal{G}^{s}=\lim_{V^{s}\rightarrow 0}\frac{\partial (I_{\uparrow
}+I_{\downarrow })}{\partial V^{s}}=\frac{e^{2}}{h}[T_{\uparrow }(\mu
)-T_{\downarrow }(\mu )],  \label{spin conductance}
\end{equation}%
which is proportional to the difference between the transmission probabilities
of two spin components. The physical picture of this definition is very
clear. Because spin $\uparrow $ and $\downarrow $ are under opposite biases,
$I_{\uparrow }$ and $I_{\downarrow }$ tend to flow along opposite
directions. $|I_{\uparrow }|$ and $|I_{\downarrow }|$ must be unequal to
generate a net charge current, which is an experimentally measurable quantity.
Note that $|I_{\uparrow }|$ can be larger or smaller than $|I_{\downarrow
}|$, depending on the ability of the dot in conducting electron with spin $%
\uparrow $ and $\downarrow $. Therefore, the spin conductance can be either
positive or negative. From Eq. (\ref{spin conductance}), one immediately
realizes that if the spin symmetry of a mesoscopic system is broken, it can
be probed by the spin conductance. Experimentally, one just applies a very
small spin bias $\Delta V^{s}$, then measures the net charge current $\Delta
I$ (note that there is no need to measure the polarization of the current;
instead, the measurement of current direction is required), and performs $%
\Delta I/\Delta V^{s}$ to have an approximation of Eq. (\ref{spin
conductance}).

\section{\label{Sec:deduction of G2}Deduction of Green's functions in Eq. (\ref{G2})}

This part of calculation is inspired by the work by Tolea and Bulka,\cite%
{Tolea2007} in which an $\langle S^{z}\rangle =0$ ($\langle s^{z}_1\rangle$
in our model) case was studied. The equation of motion for $\langle
\langle d_{2\uparrow }|d_{2\uparrow }^{\dag }\rangle \rangle $ in the
Fourier space is written as
\begin{eqnarray}
&&(\omega -\epsilon _{2}+\frac{i}{2}\Gamma _{2})\langle \langle d_{2\uparrow
}|d_{2\uparrow }^{\dag }\rangle \rangle  \notag \\
&=&1+\frac{J}{2}\langle \langle d_{2\downarrow }s_{1}^{-}|d_{2\uparrow
}^{\dag }\rangle \rangle +\frac{J}{2}\langle \langle d_{2\uparrow
}s_{1}^{z}|d_{2\uparrow }^{\dag }\rangle \rangle  \notag \\
&&+U_{2}\langle \langle d_{2\uparrow }n_{2\downarrow }|d_{2\uparrow }^{\dag
}\rangle \rangle ,
\end{eqnarray}%
where and in the following we suppress all the $r$ superscripts of the
retarded Green's functions, and introduce the notation $\langle \langle
A|B\rangle \rangle $ as the Fourier transform of $-i\theta (t)\langle
\{A(t),B\}\rangle $. Continue writing the equation of motion of Green's
functions that contain only the operators $d_{2\sigma }$, $d_{2\sigma }^{\dag }$,
and $s_{1}^{z,\pm }$ until no more new Green's function is produced,
\begin{eqnarray}
&&(\omega -\epsilon_2+\frac{J}{4})\langle \langle d_{2\downarrow
}s_{1}^{-}|d_{2\uparrow }^{\dag }\rangle \rangle  \notag \\
&=&\frac{J}{4}\langle \langle d_{2\uparrow }|d_{2\uparrow }^{\dag }\rangle
\rangle -\frac{J}{2}\langle \langle d_{2\uparrow }s_{1}^{z}|d_{2\uparrow
}^{\dag }\rangle \rangle  \notag \\
&&+J\langle \langle d_{2\uparrow }n_{2\downarrow }s_{1}^{z}|d_{2\uparrow
}^{\dag }\rangle \rangle  \notag \\
&&+(\frac{J}{2}+U_{2})\langle \langle d_{2\downarrow }n_{2\uparrow
}s_{1}^{-}|d_{2\uparrow }^{\dag }\rangle \rangle  \notag \\
&&+\sum_{k,\alpha }V_{k\alpha \downarrow }\langle \langle c_{k\alpha
\downarrow }s_{1}^{-}|d_{2\uparrow }^{\dag }\rangle \rangle ,
\end{eqnarray}

\begin{eqnarray}
&&(\omega -\epsilon_{2})\langle \langle d_{2\uparrow }s_{1}^{z}|d_{2\uparrow
}^{\dag }\rangle \rangle  \notag \\
&=&\langle s_{1}^{z}\rangle +\frac{J}{8}\langle \langle d_{2\uparrow
}|d_{2\uparrow }^{\dag }\rangle \rangle -\frac{J}{4}\langle \langle
d_{2\downarrow }s_{1}^{-}|d_{2\uparrow }^{\dag }\rangle \rangle  \notag \\
&&+\frac{J}{2}\langle \langle d_{2\downarrow }n_{2\uparrow
}s_{1}^{-}|d_{2\uparrow }^{\dag }\rangle \rangle +U_{2}\langle \langle
d_{2\uparrow }n_{2\downarrow }s_{1}^{z}|d_{2\uparrow }^{\dag }\rangle \rangle
\notag \\
&&+\sum_{k,\alpha }V_{k\alpha \uparrow }\langle \langle c_{k\alpha \uparrow
}s_{1}^{z}|d_{2\uparrow }^{\dag }\rangle \rangle ,
\end{eqnarray}%
\begin{eqnarray}
&&(\omega -\epsilon_2-U_{2})\langle \langle d_{2\uparrow }n_{2\downarrow
}|d_{2\uparrow }^{\dag }\rangle \rangle  \notag \\
&=&\langle n_{2\downarrow }\rangle +\frac{J}{2}\langle \langle
d_{2\downarrow }n_{2\uparrow }s_{1}^{-}|d_{2\uparrow }^{\dag }\rangle
\rangle +\frac{J}{2}\langle \langle d_{2\uparrow }n_{2\downarrow
}s_{1}^{z}|d_{2\uparrow }^{\dag }\rangle \rangle  \notag \\
&&+\sum_{k,\alpha }V_{k\alpha \downarrow }\langle \langle d_{2\uparrow
}d_{2\downarrow }^{\dag }c_{k\alpha \downarrow }|d_{2\uparrow }^{\dag
}\rangle \rangle  \notag \\
&&-\sum_{k,\alpha }V_{k\alpha \downarrow }\langle \langle d_{2\uparrow
}c_{k\alpha \downarrow }^{\dag }d_{2\downarrow }|d_{2\uparrow }^{\dag
}\rangle \rangle  \notag \\
&&+\sum_{k,\alpha }V_{k\alpha \uparrow }\langle \langle c_{k\alpha \uparrow
}n_{2\downarrow }|d_{2\uparrow }^{\dag }\rangle \rangle ,
\end{eqnarray}%
\begin{eqnarray}
&&(\omega -\epsilon _{2}-U_{2})\langle \langle d_{2\uparrow }n_{2\downarrow
}s_{1}^{z}|d_{2\uparrow }^{\dag }\rangle \rangle  \notag \\
&=&\langle n_{2\downarrow }s_{1}^{z}\rangle +\frac{J}{4}\langle \langle
d_{2\downarrow }n_{2\uparrow }s_{1}^{-}|d_{2\uparrow }^{\dag }\rangle
\rangle +\frac{J}{8}\langle \langle d_{2\uparrow }n_{2\downarrow
}|d_{2\uparrow }^{\dag }\rangle \rangle  \notag \\
&&+\sum_{k,\alpha }V_{k\alpha \downarrow }\langle \langle d_{2\uparrow
}d_{2\downarrow }^{\dag }c_{k\alpha \downarrow }s_{1}^{z}|d_{2\uparrow
}^{\dag }\rangle \rangle  \notag \\
&&-\sum_{k,\alpha }V_{k\alpha \downarrow }\langle \langle d_{2\uparrow
}c_{k\alpha \downarrow }^{\dag }d_{2\downarrow }s_{1}^{z}|d_{2\uparrow
}^{\dag }\rangle \rangle  \notag \\
&&+\sum_{k,\alpha }V_{k\alpha \uparrow }\langle \langle c_{k\alpha \uparrow
}n_{2\downarrow }s_{1}^{z}|d_{2\uparrow }^{\dag }\rangle \rangle ,
\end{eqnarray}%
\begin{eqnarray}
&&(\omega -\epsilon_2-U_2-\frac{J}{4})\langle \langle d_{2\downarrow
}n_{2\uparrow }s_1^-|d_{2\uparrow }^{\dag }\rangle \rangle  \notag \\
&=&-\langle d_{2\uparrow }^{\dag }d_{2\downarrow }s_1^-\rangle +\frac{J}{4}%
\langle \langle d_{2\uparrow }n_{2\downarrow }|d_{2\uparrow }^{\dag }\rangle
\rangle  \notag \\
&&+\frac{J}{2}\langle \langle d_{2\uparrow }n_{2\downarrow
}s_1^{z}|d_{1\uparrow }^{\dag }\rangle \rangle  \notag \\
&&-\sum_{k}V_{k\alpha \uparrow }\langle \langle d_{2\downarrow }c_{k\uparrow
}^{\dag }d_{2\uparrow }s_{1}^{-}|d_{2\uparrow }^{\dag }\rangle \rangle
\notag \\
&&+\sum_{k}V_{k\alpha \uparrow }\langle \langle d_{2\downarrow }d_{2\uparrow
}^{\dag }c_{k\uparrow }s_{1}^{-}|d_{2\uparrow }^{\dag }\rangle \rangle
\notag \\
&&+\sum_{k\alpha }V_{k\alpha \downarrow }\langle \langle c_{k\alpha
\downarrow }n_{2\uparrow }s_{1}^{-}|d_{2\uparrow }^{\dag }\rangle \rangle ,
\end{eqnarray}%
where we have taken advantage of the single occupation of dot 1, i.e.,
$n_{1\uparrow }+n_{1\downarrow }=1$, so that
\begin{eqnarray}
&&s_{1}^{z/+}s_{1}^{+/z}=\pm \frac{1}{2}s_{2}^{+},\
s_{1}^{z/-}s_{1}^{-/z}=\mp \frac{1}{2}s_{2}^{-},  \notag \\
&&s_{1}^{\pm }s_{1}^{\mp }=\frac{1}{2}\pm s_{1}^{z},\ (s_{1}^{z})^{2}=\frac{1%
}{4}.
\end{eqnarray}%
Using the approximation scheme proposed by the previous authors to treat
quantum dots weakly coupled to electrodes,\cite{ Tolea2007,Bulka2004}
\begin{eqnarray}
\sum_{k\alpha }V_{k\alpha \uparrow }\langle \langle c_{k\alpha \uparrow
}s_{1}^{z}|d_{2\uparrow }^{\dag }\rangle \rangle &\approx &-\frac{i}{2}%
\Gamma _{2}\langle \langle d_{2\uparrow }s_{1}^{z}|d_{2\uparrow }^{\dag
}\rangle \rangle ,  \notag \\
\sum_{k\alpha }V_{k\alpha \downarrow }\langle \langle c_{k\alpha \downarrow
}s_{1}^{-}|d_{2\uparrow }^{\dag }\rangle \rangle &\approx &-\frac{i}{2}%
\Gamma _{2}\langle \langle d_{2\downarrow }s_{1}^{-}|d_{2\uparrow }^{\dag
}\rangle \rangle ,  \notag \\
\sum_{k\alpha }V_{k\alpha \uparrow }\langle \langle c_{k\alpha \uparrow
}n_{2\downarrow }|d_{2\uparrow }^{\dag }\rangle \rangle &\approx &-\frac{i}{2%
}\Gamma _{2}\langle \langle d_{2\uparrow }n_{2\downarrow }|d_{2\uparrow
}^{\dag }\rangle \rangle ,  \notag \\
\sum_{k\alpha }V_{k\alpha \uparrow }\langle \langle c_{k\alpha \uparrow
}n_{2\downarrow }s_{1}^{z}|d_{2\uparrow }^{\dag }\rangle \rangle &\approx &-%
\frac{i}{2}\Gamma _{2}\langle \langle d_{2\uparrow }n_{2\downarrow
}s_{1}^{z}|d_{2\uparrow }^{\dag }\rangle \rangle ,  \notag \\
\sum_{k\alpha }V_{k\alpha \downarrow }\langle \langle c_{k\alpha \downarrow
}n_{2\uparrow }s_{1}^{-}|d_{2\uparrow }^{\dag }\rangle \rangle &\approx &-%
\frac{i}{2}\Gamma _{2}\langle \langle d_{2\downarrow }n_{2\uparrow
}s_{1}^{-}|d_{2\uparrow }^{\dag }\rangle \rangle ,  \notag \\
&&
\end{eqnarray}%
where $\Gamma _{2}=\Gamma _{2}^{L}+\Gamma _{2}^{R}$. Moreover,
simultaneous hopping in and out of the quantum dot are regarded to cancel each other,\cite{Bulka2004}
\begin{eqnarray}
\langle \langle d_{2\uparrow }c_{k\alpha \downarrow }^{\dag }d_{2\downarrow
}|d_{2\uparrow }^{\dag }\rangle \rangle &\approx&\langle \langle
d_{2\uparrow }d_{2\downarrow }^{\dag }c_{k\alpha \downarrow }|d_{2\uparrow
}^{\dag }\rangle \rangle ,  \notag \\
\langle \langle d_{2\uparrow }c_{k\alpha \downarrow }^{\dag }d_{2\downarrow
}s_{1}^{z}|d_{2\uparrow }^{\dag }\rangle \rangle &\approx&\langle \langle
d_{2\uparrow }d_{2\downarrow }^{\dag }c_{k\alpha \downarrow
}s_{1}^{z}|d_{2\uparrow }^{\dag }\rangle \rangle ,  \notag \\
\langle \langle d_{2\downarrow }c_{k\alpha\uparrow }^{\dag }d_{2\uparrow
}s_{1}^{-}|d_{2\uparrow }^{\dag }\rangle \rangle &\approx&\langle \langle
d_{2\downarrow }d_{2\uparrow }^{\dag }c_{k\alpha\uparrow
}s_{1}^{-}|d_{2\uparrow }^{\dag }\rangle \rangle .
\end{eqnarray}%
The above approximations are valid only when the quantum dot is weakly
coupled to the leads and for temperatures higher than the Kondo temperature.
The advantage of the approximation is that it retains the full inter-dot
correlations and gives a correct physical picture in the Coulomb blockade
regime. After applying the truncation approximation, the equation of motion
for the spin $\uparrow $ retarded Green's function of dot 2 in the
singly occupied regime of dot 1 can be obtained as Eq. (\ref{G2}). The
equation of motion for $\langle \langle d_{2\downarrow }|d_{2\downarrow
}^{\dag }\rangle \rangle $ can be obtained similarly.

\end{document}